\documentclass{jacs}
\usepackage{nomencl}

\author{Alec Glisman}
\affiliation{Division of Chemistry and Chemical Engineering, California Institute of Technology, Pasadena, California 91125, USA}
\altaffiliation{Contributed equally to this work}

\author{Sriteja Mantha}
\affiliation{Division of Chemistry and Chemical Engineering, California Institute of Technology, Pasadena, California 91125, USA}
\altaffiliation{Contributed equally to this work}

\author{Decai Yu}
\affiliation{The Dow Chemical Company, Core R\&D, 633 Washington St., Midland, Michigan 48674, USA}

\author{Eric Wasserman}
\affiliation{The Dow Chemical Company, Consumer Solutions R\&D, 400 Arcola Road, Collegeville, Pennsylvania 19426, USA}

\author{Scott Backer}
\affiliation{The Dow Chemical Company, Consumer Solutions R\&D, 400 Arcola Road, Collegeville, Pennsylvania 19426, USA}

\author{Zhen-Gang Wang}
\affiliation{Division of Chemistry and Chemical Engineering, California Institute of Technology, Pasadena, California 91125, USA}
\email{zgw@caltech.edu}

\title{Multi-valent Ion Mediated Polyelectrolyte Association and Structure}


\begin{document}

\begin{abstract}
    \begingroup
    Polyelectrolytes are commonly used to chelate multi-valent ions in aqueous solutions, playing a critical role in water softening and the prevention of mineralization.
At sufficient ionic strength, ion-mediated polyelectrolyte--polyelectrolyte interactions can precipitate polyelectrolyte--ion complexes, a phenomenon known as ``like-charge attraction''.
While the significant influence of small ions on polyelectrolyte solution phase behavior is recognized, the precise molecular mechanisms driving the counterintuitive phenomenon remain largely elusive.
In this study, we employ all-atom molecular dynamics simulations to investigate the molecular mechanism of like-charge attraction between two poly(acrylic acid) (PAA) chains in solution.
We find that moderate quantities of Ca$^{2+}$ ions induce attraction between PAA chains, facilitated by the formation of PAA--Ca$^{2+}$--PAA bridges and a significant increase in the coordination of Ca$^{2+}$ ions by the PAA chains.
At high Ca$^{2+}$ number densities, ion bridges are disfavored due to electrostatic screening, yet the chains are still attracted to each other due to solvent-mediated interactions between the chains and their chelated ions.
The insights gleaned from this study not only enrich our understanding of the intricate mechanism of like-charge attraction between polyanions in solution but also illuminate the influence of multi-valent ions on polyelectrolyte interactions.

    \endgroup
\end{abstract}

\begingroup
\section{Introduction}\label{sec:intro}

Aqueous polyelectrolyte (PE) solutions find widespread utility in diverse fields, including water treatment \cite{bolto2007organic}, drug delivery \cite{lankalapalli2009polyelectrolyte}, and scale (CaCO$_3$) inhibition \cite{reddy_calcite_2001}, among others.
\nomenclature{PE}{Polyelectrolyte}
For many of these applications, it is crucial to control the behavior of the polyelectrolyte in the solution phase.
By carefully tuning the ionic strength, intriguing phenomena known as ``salting out'' and ``salting in'' can be induced, leading to the precipitation or re-dissolution of polyelectrolytes, respectively \cite{arakawa1984mechanism,allahyarov_attraction_1998,curtis_protein-protein_2002,saluja_anion_2009,li_strong_2017}.
Therefore, a detailed understanding of the relationship between polyelectrolyte structure and PE--ion interactions is essential for the systematic design of advanced polyelectrolyte materials and additives.

Polyelectrolytes, such as poly(acrylic acid) (PAA), are commonly used as scale inhibitors due to their ability to chelate Ca$^{2+}$ ions \cite{sinn_isothermal_2004,gindele_binding_2022} as well as modify CaCO$_3$ crystal growth \cite{reddy_calcite_2001,yu_effects_2004,jada_effect_2007,aschauer_growth_2010}.
\nomenclature{PAA}{Poly(acrylic acid)}
The effectiveness of PAA, and other scale inhibitors, relies heavily on the polymer's ability to chelate many Ca$^{2+}$ while preventing precipitation of the polyelectrolyte and chelated ion complex.
Past experimental studies, such as those by Huber \cite{huber1993calcium,schweins2001collapse}, have demonstrated that the addition of Ca$^{2+}$ induces a transition in polymer conformation from an extended coil to a collapsed chain, which results in the precipitation of PE--ion complexes once the polymer's binding capacity is exceeded.
Subsequent work conducted by Sinn et al. \cite{sinn_isothermal_2004} found that a screened-Coulomb ion-polyelectrolyte interaction model could not explain the observed precipitation phenomenon, suggesting a more detailed understanding of Ca$^{2+}$ mediated interactions is necessary.

Polyelectrolyte theory suggests that a polyelectrolyte's behavior and adsorption properties in solution are strongly tied to its chain conformation, which in turn is influenced by solution ionic strength and ionic valency \cite{de1995precipitation,castelnovo_charge_2000,huang2002polyelectrolytes,liu2003polyelectrolyte,kundagrami2008theory,lee2009phase,muthukumar_50th_2017}.
Coarse-grained theoretical models have captured the addition of multivalent ions leading to chain collapse.
However, precipitation has been attributed to differing mechanisms, including correlations between chelated ions, ion bridging between chains, or charge neutralization of the polymer in various studies \cite{ha_counterion-mediated_1997,arenzon_simple_1999,wittmer_precipitation_1995,solis2000collapse}.
While theory and experiments predict precipitation of polyelectrolyte--Ca$^{2+}$ complexes at certain conditions, to design novel polyelectrolytes that stay soluble in aqueous solution, we need to understand the molecular principles that govern the Ca$^{2+}$ induced association between like-charged polyelectrolytes.
Molecular dynamics (MD) simulations provide a framework for such an investigation.
\nomenclature{MD}{Molecular Dynamics}

Several seminal MD studies have investigated the behavior of polyelectrolytes in aqueous CaCl$_2$ solutions.
The work of Molnar and Rieger \cite{molnar_like-charge_2005} provided evidence for the attraction of polyanions in solution by showing that two PAA chains in a solution were more prone to association as the number of Ca$^{2+}$ ions increased.
However, the limited simulation box size (6 nm) and integration time (10 ns) precluded observation of different ion binding environments, polymer conformations, and transitions between states.
Subsequent single-chain PAA simulations by the Parrinello group \cite{bulo_site_2007,tribello_binding_2009} corroborated the stability of ion bridging and hypothesized that interchain ion bridges were responsible for the observed attraction between chains.

While these prior studies have yielded valuable insights, our preceding work \cite{mantha2023adsorption} underscored a critical limitation: using classical force fields to model electrostatic interactions involving Ca$^{2+}$ without accounting for polarization effects results in exceedingly long Ca$^{2+}$-PAA relaxation times.
Such models can erroneously predict charge inversion of PAA-Ca$^{2+}$ complexes, a phenomena that is not supported by experimental data \cite{sinn_isothermal_2004,gindele_binding_2022}.
This discrepancy implies that earlier MD investigations concerning Ca$^{2+}$ mediated like-charge interactions could suffer from insufficient sampling of PAA-Ca$^{2+}$ populations and polyelectrolyte conformations.

To overcome these challenges, our previous work \cite{mantha2023adsorption} utilized the Electronic Continuum Correction (ECC) method, modeling the electronic polarization effects in a mean-field manner.
This approach enabled us to mirror the polyelectrolyte--ion binding capacities observed in experiments \cite{sinn_isothermal_2004,gindele_binding_2022}.
\nomenclature{ECC}{Electronic Continuum Correction}
We also calculated the adsorption isotherm for Ca$^{2+}$ ions on a single PAA chain.
The findings showed a transition from an extended coil to a collapsed chain conformation as the number of Ca$^{2+}$ ions increased, a result consistent with experimental observations \cite{huber1993calcium,gindele_binding_2022}.
We attributed this transition to the formation of intrachain ion bridges which caused a chain collapse.
Despite these advancements, there remain unresolved questions.
For example, it is unclear how Ca$^{2+}$ ions mediate interchain interactions, and whether the formation of interchain ion bridges constitutes the dominant mechanism driving like-charge attraction between PAA chains in solution.

In the present study, we use our previously developed all-atom MD model \cite{mantha2023adsorption} to investigate the molecular mechanism of like-charge attraction between two PAA chains in solution.
Our enhanced sampling protocol provides an efficient exploration of the polyelectrolyte conformational space and ion binding environments, enabling the identification of the dominant conformations of PAA chains in solution.
We find that at moderate Ca$^{2+}$ numbers, the PAA chains are attracted to each other due to the formation of direct PAA--Ca$^{2+}$--PAA bridges between the chains as well as a significant increase in the number of ion and polymer contacts.
However, at high Ca$^{2+}$ numbers, direct ion bridges between chains are disfavored due to electrostatic screening, yet the chains are still attracted to each other mainly due to solvent-mediated interactions between chelated ions on one chain and the carboxylate groups on the other chain. 
We then compute the precipitation concentration of PAA in solution using the second osmotic virial coefficient to quantify the net attraction between PAA chains.
To analyze the dominant conformations of PAA chains interacting in solution and to investigate potential transition pathways between these conformations, we employ machine learning techniques.

The rest of the manuscript is organized as follows.
We first describe our models and specific parameters used in the enhanced sampling algorithms for the molecular dynamics simulations.
Next, we report the numerical data obtained from our calculations and discuss their implications.
We then conclude the article with an outlook on the path forward for molecular design of antiscalant polyelectrolytes.

\section{Methods}\label{sec:methods}

\paragraph{Molecular Dynamics Simulations.}

Molecular dynamics simulations were performed using GROMACS (version 2022.3) MD engine \cite{berendsen1995gromacs,van2005gromacs,abraham2015gromacs} integrated with PLUMED (version 2.8.1) \cite{bonomi2009plumed,tribello2014plumed,bonomi2019promoting}.
The PAA chains were modeled as atactic and fully deprotonated (pH $\geq 11$) with Na$^{+}$ counter-ions, and constructed using CHARMM-GUI \cite{jo2008charmm,choi2021charmm} with 16 monomers per chain.
The PAA chains were then solvated in a 12 nm cubic box of SPC/E water \cite{berendsen1987missing} with Packmol \cite{martinez2009packmol}.

The general AMBER force field (GAFF) \cite{wang2004development,wang2006automatic} was employed to model the PAA chains, following the protocol used in our single-chain PAA studies \cite{mantha2023adsorption} and originally validated by Mintis and Mavrantzas \cite{mintis_effect_2019}.
\nomenclature{GAFF}{General AMBER Force Field}
As electrostatic interactions are known to be overly strong in non-polarizable force fields \cite{martinek_calcium_2018}, we utilized the electronic continuum correction (ECC) method \cite{leontyev_electronic_2009,duboue-dijon_practical_2020} to more accurately describe the electrostatic interactions between PAA and Ca$^{2+}$ ions.
\nomenclature{ECC}{Electronic Continuum Correction}
The ion parameters for Ca$^{2+}$ and Cl$^{-}$ were taken from the electronic continuum correction with rescaling (ECCR) parameters of Martinek et al. \cite{martinek_calcium_2018}, while those for Na$^{+}$ were taken from the ECCR optimized parameters of Kohagen et al. \cite{kohagen_accurate_2014}.
\nomenclature{ECCR}{Electronic Continuum Correction with Rescaling}
It has been shown that scaling polyelectrolyte and small ion charges is required to reproduce experimental binding results accurately \cite{duboue-dijon_binding_2018,mantha2023adsorption}.
As a result, we applied the ECC method to all PAA partial charges.

For the van der Waals interactions, a cutoff of 1.2 nm was chosen.
The long-range electrostatic interactions were calculated via the PME method \cite{darden1993particle,essmann1995smooth} with a real space cutoff of 1.2 nm.
The LINCS algorithm \cite{hess1997lincs} was utilized to constrain all bonds with hydrogen atoms, and the system equations of motion were integrated using the leap-frog algorithm.

After solvation and ion addition, the system was energy-minimized using the steepest descent algorithm for approximately 100,000 steps.
The system was then equilibrated in the NVT ensemble at 300 K for 10 ns with 1 fs time steps, using the Nos\'{e}--Hoover thermostat \cite{nose1984unified,hoover1985canonical} with a 0.25 ps relaxation time constant.
\nomenclature{NVT}{Constant Number of Particles, Volume, and Temperature}
Further equilibration was carried out for another 10 ns in the NPT ensemble at 300 K and 1 bar using the Parrinello--Rahman barostat \cite{parrinello1981polymorphic} with a 5 ps relaxation time constant.
\nomenclature{NPT}{Constant Number of Particles, Pressure, and Temperature}

The production phase simulations were conducted in the NVT ensemble at 300 K using a time step of 2 fs.
The sampling of the equilibrium ensemble was enhanced by applying a combined well-tempered metadynamics \cite{laio_escaping_2002,barducci_well-tempered_2008,barducci_metadynamics_2011} and HREMD \cite{wang_replica_2011,bussi_hamiltonian_2014} protocol.
The well-tempered metadynamics protocol was used for improved sampling of the two-chain distance free energy surface (FES) \cite{barducci_well-tempered_2008,barducci_metadynamics_2011}.
Though metadynamics sampled the two-chain distance effectively, the slow relaxation of the polymer conformations and ion binding environments prevented adequate convergence of the FES.
Consequently, we employed the HREMD protocol to efficiently sample the polymer conformational space \cite{park_atomistic_2012} and ion binding environments \cite{robbins_effect_2013}.
\nomenclature{HREMD}{Hamiltonian Replica Exchange Molecular Dynamics}
Concurrently, the well-tempered metadynamics protocol was used for improved sampling of the two-chain distance free energy surface (FES) \cite{barducci_well-tempered_2008,barducci_metadynamics_2011}.
\nomenclature{FES}{Free Energy Surface}

Coordinate exchanges between replicas were attempted every 100 steps, and the number of replicas was set to ensure an average exchange acceptance rate of approximately 25\%.
The systems with 0, 8, 16, and 32 Ca$^{2+}$ had 16 replicas, while those with 64 and 128 Ca$^{2+}$ had 24 replicas.
In each replica, the electrostatic, Lennard-Jones, and dihedral interactions for all solute atoms were scaled by a parameter $\lambda$, ranging from 1 (unbiased) and 0.68 (most biased) geometrically distributed to improve exchange acceptance rates \cite{bussi_hamiltonian_2014}.

Each replica contained an independent well-tempered metadynamics bias potential with an initial Gaussian height of 1.2 kJ/mol, Gaussian width of 0.025 nm, deposition stride of 500 steps, and bias factor of 10.
The metadynamics biases of each replica were evaluated in the Metropolis coordinate exchange probability.
The collective variable (CV) in metadynamics was the distance between the centers of mass of the two PAA chains, facilitating the sampling of various associated and dissociated states.
\nomenclature{CV}{Collective Variable}
An additional harmonic wall was placed at 6 nm to prevent sampling beyond the minimum image.
Replicas were equilibrated for 25 ns at 300 K before production sampling, and the metadynamics bias was added.
Each replica simulation was run for at least 250 ns, yielding a total production simulation time of 29.4 $\mu$s.

The reported data in our study were collected from the $\lambda = 1$ replica, and statistics were calculated by reweighting this data based on the Boltzmann factor of the metadynamics and harmonic wall biases.
Uncertainties, when reported, were estimated using block averaging on correlated MD data to obtain the standard error of the sample mean, which was then converted to 95\% confidence intervals.
Analysis of the simulation data involved the use of MDAnalysis \cite{michaud2011mdanalysis,gowers2016mdanalysis} and custom Python scripts, while VMD \cite{HUMP96,STON1998} was employed for visualizing the generated trajectories.
\nomenclature{VMD}{Visual Molecular Dynamics}

\paragraph{Stability of Polyelectrolyte Solution.}

Determination of the solution phase boundary from molecular simulation is non-trivial, as phase transitions are macroscopic events, and the range of system configurational space is large.
This boundary typically comprises a polymer-dilute branch and a polymer-rich branch interconnected at the critical concentration,
and as shown below, the spinodal of the dilute branch can be estimated with a leading-order expansion of the osmotic virial equation of state.
In contrast, the branch corresponding to higher polymer concentrations would require many chain simulations and the enhanced sampling approaches detailed above would become computationally prohibitive.
Despite the difference in polymer concentrations and the complexities involved in their prediction, one can anticipate similar key physics and interactions on both branches of the spinodal.
This is due to the fact that precipitation is primarily driven by local interactions between the polymer chains and the ions.
Thus, regardless of the concentration, the essential characteristics of these interactions remain the same.

To estimate the spinodal from the simulation data, the primary property of interest from the simulations is the potential of mean force (PMF) between the two PAA chains as a function of the center of mass separation distance.
\nomenclature{PMF}{Potential of Mean Force}
The PMF provides a local measure of the relative stability of associated and dissociated chain configurations.
If the PMF at narrow separations is negative, the chains favor association, which may lead to precipitation of the polyelectrolyte at sufficiently high concentrations.

Following previous studies \cite{neal_molecular_1998,lund_mesoscopic_2003,stark_toward_2013}, we can then calculate the second osmotic virial coefficient as
\begin{equation}\label{eq:second_virial_coefficient}
    B_2{(T)}
    = - 2 \pi
    \int_{0}^{\infty} dr \, r^2
    \left(
    e^{- \beta U_\mathrm{PMF}{(r)}}
    - 1
    \right)
    ~,
\end{equation}
where $r$ denotes the center of mass distance between the two PAA chains, $\beta = 1 / k_B T$, and $U_\mathrm{PMF}{(r)}$ is the PMF obtained from the simulation.
Rigorously, the PMF should be calculated at fixed chemical potential of the ions, but constant chemical potential simulations are computationally prohibitive for the systems studied here.
However, the PMF calculated at fixed number density of ions (canonical ensemble) has been shown to reproduce experimental trends in phase behavior for similar macromolecules with added salt \cite{lund_mesoscopic_2003}.

A positive $B_2{(T)}$ indicates a repulsive interaction, while a negative $B_2{(T)}$ indicates an attractive interaction.
To calculate the maximum polyelectrolyte density of the suspension before the onset of polyelectrolyte--ion precipitation, we utilized the leading-order term in the virial expansion of the osmotic pressure \cite{neal1998molecular,hill2013statistical} to find the spinodal of the polyelectrolyte solution as
\begin{equation}\label{eq:critical_pe_concentration}
    \rho_\mathrm{PE} = - \frac{1}{2 \, B_2{(T)}}
    \,\, \mid \,\,
    B_2{(T)} < 0
    ~.
\end{equation}
The spinodal polyelectrolyte chain concentration allows us to quantitatively determine the effects of Ca$^{2+}$ ions on the polyelectrolyte solution phase behavior and evaluate the emergence of like-charge attraction.

\paragraph{Autoencoder Neural Network.}

We employed machine learning models to analyze the large amount of simulation data generated and to map the polyelectrolyte conformational space in a low-dimensional representation.
Autoencoder (AE) networks, a type of neural network, have emerged as powerful tools in understanding and predicting complex systems.
\nomenclature{AE}{Autoencoder}
Recent studies have shown successful applications of AE networks in mapping the conformational space of proteins and oligomers \cite{lemke_encodermap_2019,lemke_encodermapii_2019,bandyopadhyay_deep_2021,glielmo_unsupervised_2021}.
An AE network consists of an encoder that maps the input to a lower-dimensional representation and a decoder that reconstructs the original input from the lower-dimensional representation.
Both networks are symmetric, containing the same number of layers and neurons.
The encoded lower-dimensional representation, known as the latent space, provides a compressed representation of the input data and aids in visualizing the phase space.

In line with the approach proposed by Bandyopadhyay and Mondal \cite{bandyopadhyay_deep_2021}, we trained the AE network using the pairwise distances between the C$_\alpha$ atoms of the PAA backbone.
The encoder was built with three hidden layers comprising 496, 128, and 32 neurons, fully connected with a latent space output of 2 dimensions.
The training was conducted over 150 epochs using the Adam optimizer with a learning rate of 0.001 and batches of 256 randomly chosen conformations.
We used the mean squared error between the input and output as the loss function and applied an $L_2$ penalty of 0.00001 for weight regularization.
Furthermore, to test the model, 20\% of the data was withheld.
All aspects of model development and training were facilitated by the PyTorch library \cite{paszke2017automatic,paszke2019pytorch}.

\section{Results and Discussion}\label{sec:results}

The 16-mer PAA two-chain PMF curves for each number of Ca$^{2+}$ ions ($N_{\mathrm{Ca}^{2+}}$) are shown in Figure~\ref{fig:paa_homopolymer_interchain_pmf}.
The reference state for each system is the free energy at $r = 4$ nm, where the average two-chain interactions have plateaued for all calcium-containing systems.
In the absence of Ca$^{2+}$ ions, the system has repulsive (${\Delta F > 0}$) interactions at all distances, as expected.
\begin{figure}[htbp]
    \centering
    \includegraphics[width=0.95\linewidth]{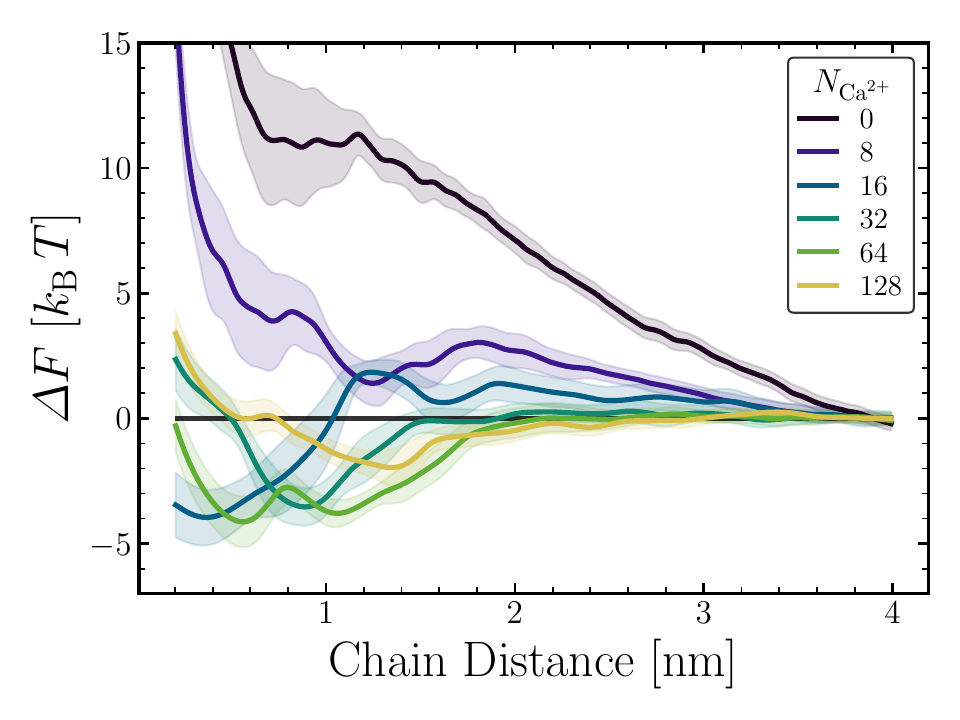}
    \caption{
        Potential of mean force for two 16-mer PAA chains with varying number of Ca$^{2+}$ ions.
    }
    \label{fig:paa_homopolymer_interchain_pmf}
\end{figure}
At small numbers of Ca$^{2+}$ ions (${N_{\mathrm{Ca}^{2+}} \approx 8}$), $\mathrm{Ca}^{2+}$ adsorption reduces the long-ranged chain repulsion and a metastable association well emerges at ${\sim}1.2$ nm.
As $N_{\mathrm{Ca}^{2+}}$ increases, the PMF develops a globally stable well at short distances ($r \leq 1.8$ nm) with a large barrier separating the associated and dissociated states.
Higher $N_{\mathrm{Ca}^{2+}}$ shifts the PMF well to larger distances and the barrier height vanishes.
Increased electrostatic screening at higher $N_{\mathrm{Ca}^{2+}}$ reduces the long-ranged repulsion between the chains, and likely contributes to the observed reduction in the barrier height.

We next estimated the precipitation conditions of the polymer--ion complex using the osmotic virial equation of state.
The second osmotic virial coefficient ($B_2$) physically represents the two-body interaction between polymer chains such that a positive $B_2$ value indicates repulsion, and a negative $B_2$ value indicates attraction.
We calculate $B_2$ using Eq.~\ref{eq:second_virial_coefficient} with the approximation that the infinite integral domain can be replaced with a finite domain of $r \in [0, 6]$ nm, which is appropriate given the short-ranged interactions and the PMF curves plateau, which causes the integrand to vanish.
\begin{figure}[htbp]
    \centering
    \includegraphics[width=0.95\linewidth]{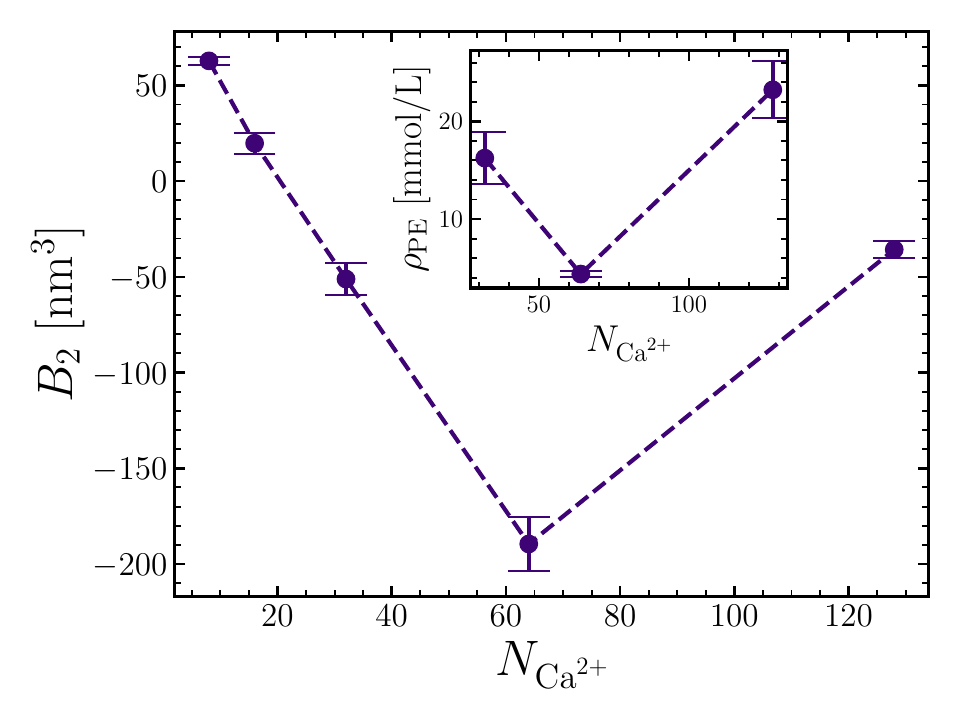}
    \caption{
        PAA second osmotic virial coefficient as a function of Ca$^{2+}$ ions in the system.
        Inset shows the corresponding polymer concentration at which the polymer--ion complex is predicted to precipitate.
        The dotted lines are guides to the eye.
    }
    \label{fig:paa_homopolymer_b2_vs_nca}
\end{figure}
We do not include the ${N_{\mathrm{Ca}^{2+}} = 0}$ system in the calculation of $B_2$ because the PMF is repulsive at all distances.
For systems with net-attraction ($B_2 < 0$), the corresponding maximum polymer concentration at which the polymer--ion complex is predicted to precipitate is calculated using Eq.~\ref{eq:critical_pe_concentration} and shown in the inset of Figure~\ref{fig:paa_homopolymer_b2_vs_nca}.
Interestingly, the system with 16 Ca$^{2+}$ ions exhibits a positive $B_2$ value, suggesting no precipitation of the polymer--ion complex (to leading order), despite possessing an attractive PMF well of ${\sim}4 k_\mathrm{B} \, T$.
We additionally observe a non-monotonic trend in the second osmotic virial coefficient.
This trend results in the spinodal polymer concentration being higher for 128 Ca$^{2+}$ than 64 Ca$^{2+}$, which implies a salting-in effect.

\begin{figure}[htbp]
    \centering
    \includegraphics[width=0.95\linewidth]{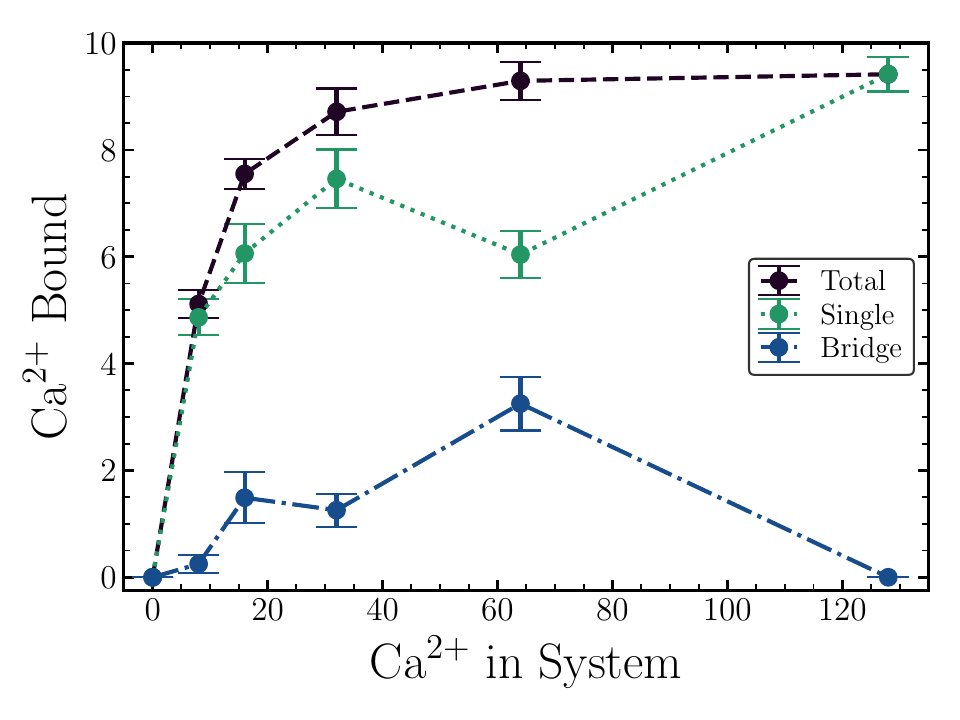}
    \caption{
        Number of calcium ions bound to PAA carboxylate groups as a function of the number of calcium ions in the system.
        Data are plotted within 1 $k_\mathrm{B} \, T$ of the minimum in the interchain potential of mean force, ensuring the relevance to the system's most stable configurations.
        Calcium ions bound to a single chain are depicted in green, while bridging ions between chains are shown in blue.
        The dotted lines are guides to the eye.
    }
    \label{fig:paa_homopolymer_nca_bound}
\end{figure}
The non-monotonic behavior of the second osmotic virial coefficient with increasing $N_{\mathrm{Ca}^{2+}}$ qualitatively agrees with theoretical predictions of salting-in effects reported by Wittmer et al. \cite{wittmer_precipitation_1995} at higher ionic strengths.
However, unlike the theoretical prediction of polymer charge inversion via ion chelation, our simulations did not show such behavior.
Instead, the number of calcium ions bound to the polymer increased and saturated below the 16 Ca$^{2+}$ count needed for charge neutrality (see Figure~\ref{fig:paa_homopolymer_nca_bound}).
The number of calcium ions bound to the two 16-mer PAA chains saturates around 9 ions, which is not enough to neutralize, much less invert, the charge of the polymer.
These results are consistent with our single-chain 32-mer PAA simulations \cite{mantha2023adsorption}, which showed that the number of calcium ions bound to the polymer saturates at approximately 0.4 ions per monomer.
Experimental studies on longer chains have similarly measured a binding capacity of about 0.3 Ca$^{2+}$ per monomer \cite{sinn_isothermal_2004,gindele_binding_2022}.

At high $N_{\mathrm{Ca}^{2+}}$, the electrostatic screening reduces the formation of ion bridges between chains, leading to a slightly weaker and longer-ranged attraction, which is consistent with our single-chain PAA simulations \cite{mantha2023adsorption} and de la Cruz \textit{et al}\cite{de1995precipitation}.
This phenomenon is the source of the salting-in behavior.
The average number of bridging Ca$^{2+}$ ions approaches zero at 128 Ca$^{2+}$ ions, yet the chains are still attracted to each other, as seen in the PMF curves (Figure~\ref{fig:paa_homopolymer_interchain_pmf}).
However, the chain-associated state remains favorable due to the presence of 9 single-chain adsorbed Ca$^{2+}$ ions.
These single-chain adsorbed Ca$^{2+}$ are often chelated by a single carboxylate group, which locally inverts the effective monomer charge from $-1$ to $+1$.
The positive charge of the effectively monovalent cation monomer screens the long-ranged chain repulsive interactions and promotes chain association.

\begin{figure*}[htb]
    \centering
    \begin{subfigure}{0.495\linewidth}
        \centering
        \captionsetup{justification=centering}
        \caption*{\large 32 Ca$^{2+}$}
        \vspace*{-1em}
        \label{sfig:paa_homopolymer_32Ca_2D_fes_distance_cn}
        \includegraphics[width=\linewidth]{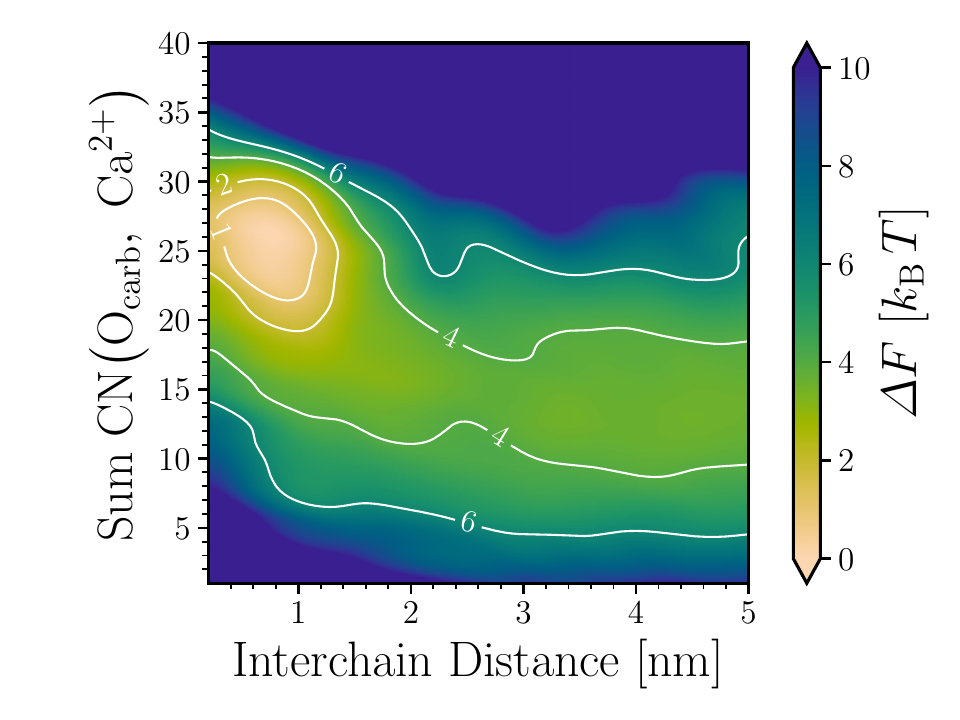}
    \end{subfigure}
    \begin{subfigure}{0.495\linewidth}
        \centering
        \captionsetup{justification=centering}
        \caption*{\large 128 Ca$^{2+}$}
        \vspace*{-1em}
        \label{sfig:paa_homopolymer_128Ca_2D_fes_distance_cn}
        \includegraphics[width=\linewidth]{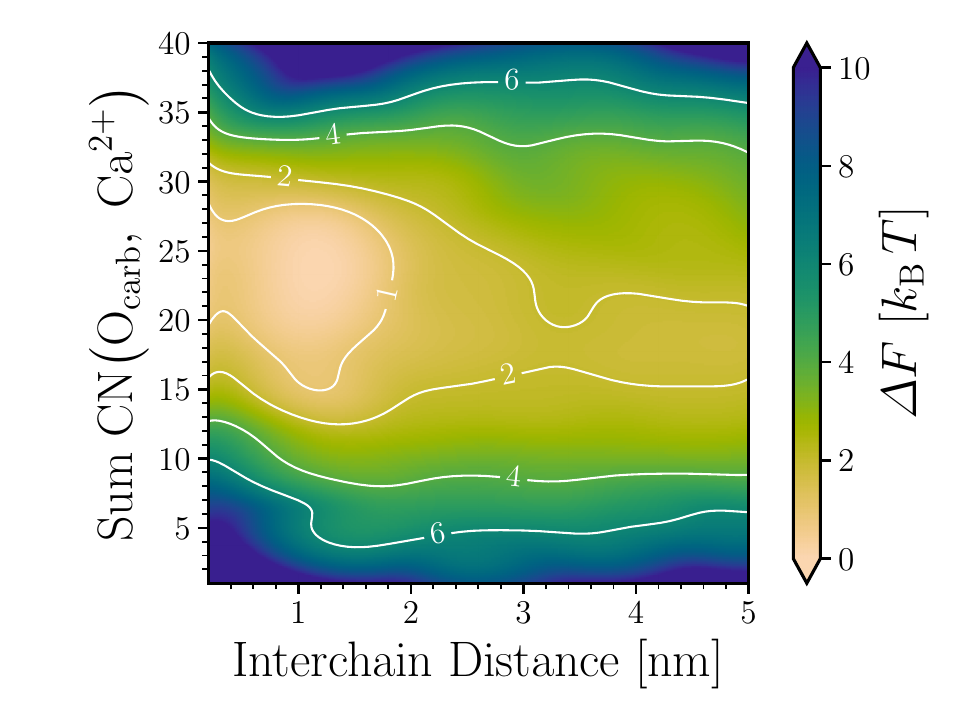}
    \end{subfigure}
    \caption{
        Two-dimensional free energy surfaces for two 16-mer PAA chains with 32 (\textbf{left panel}) and 128 (\textbf{right panel}) Ca$^{2+}$ ions.
        The horizontal axis represents the center of mass distance between the PAA chains, while the vertical axis denotes the summed coordination number of carboxylate oxygen atoms about Ca$^{2+}$ ions.
        Isolines are drawn at 1, 2, 4, and 6 $k_\mathrm{B} \, T$.
    }
    \label{fig:paa_homopolymer_2D_fes}
\end{figure*}
The depiction of Ca$^{2+}$ adsorption in Figure~\ref{fig:paa_homopolymer_nca_bound} elucidates how Ca$^{2+}$ facilitates the association of PAA chains.
However, this representation only shows the structure of associated states and not the overall driving forces propelling the chain association.
Our single-chain studies \cite{mantha2023adsorption} have shown that isolated chains similarly adsorb Ca$^{2+}$ and exhibit intrachain ion bridging.
We observe that as the two chains approach one another, the calcium-mediated interactions become more pronounced through both an increase in ion bridging and calcium adsorption, as shown in Figure.~\ref{fig:paa_homopolymer_2D_fes}.

The above trend can be quantified by evaluating the number of contacts between carboxylate groups and Ca$^{2+}$ ions.
We establish a `contact' when the distance between a carboxylate group and a Ca$^{2+}$ ion falls below 0.35 nm.
This cut-off was determined by the location of the first minima in the radial distribution function for Ca$^{2+}$ and carboxylate oxygen atoms.
The proximity denotes direct PE--ion interactions, which eliminate solvent-mediated interactions.

Figure~\ref{fig:paa_homopolymer_2D_fes} shows the free energy landscape of PE--ion contacts as a function of the interchain distance for systems with 32 and 128 Ca$^{2+}$ ions, respectively.
We focus on these two systems, as 32 Ca$^{2+}$ ions is the minimum number of ions in this study that leads to $B_2 < 0$ and 128 Ca$^{2+}$ ions exhibits non-monotonicity in the $B_2$ curve (Figure~\ref{fig:paa_homopolymer_b2_vs_nca}).
For 32 Ca$^{2+}$ ions, the stable associated states have a larger number of PE--ion contacts than the dissociated states.
As the chains approach each other, more Ca$^{2+}$ ions adsorb onto the chains due to the higher density of carboxylate groups.
This not only decreases the electrostatic repulsion between the chains but also aids in forming ion bridges.
However, for 128 Ca$^{2+}$ ions, the chains are already saturated with Ca$^{2+}$ ions in the dissociated states, and so the relative increase in the number of PE--ion contacts is less with further decrease in the interchain distance.
In addition, the overall free energy surface (FES) valley is shallower and shifted to larger chain center of mass distances, which is consistent with the decreased number of bridging Ca$^{2+}$ ions (Figure~\ref{fig:paa_homopolymer_nca_bound}).

\begin{figure*}[htb]
    \centering
    \includegraphics[width=0.80\textwidth]{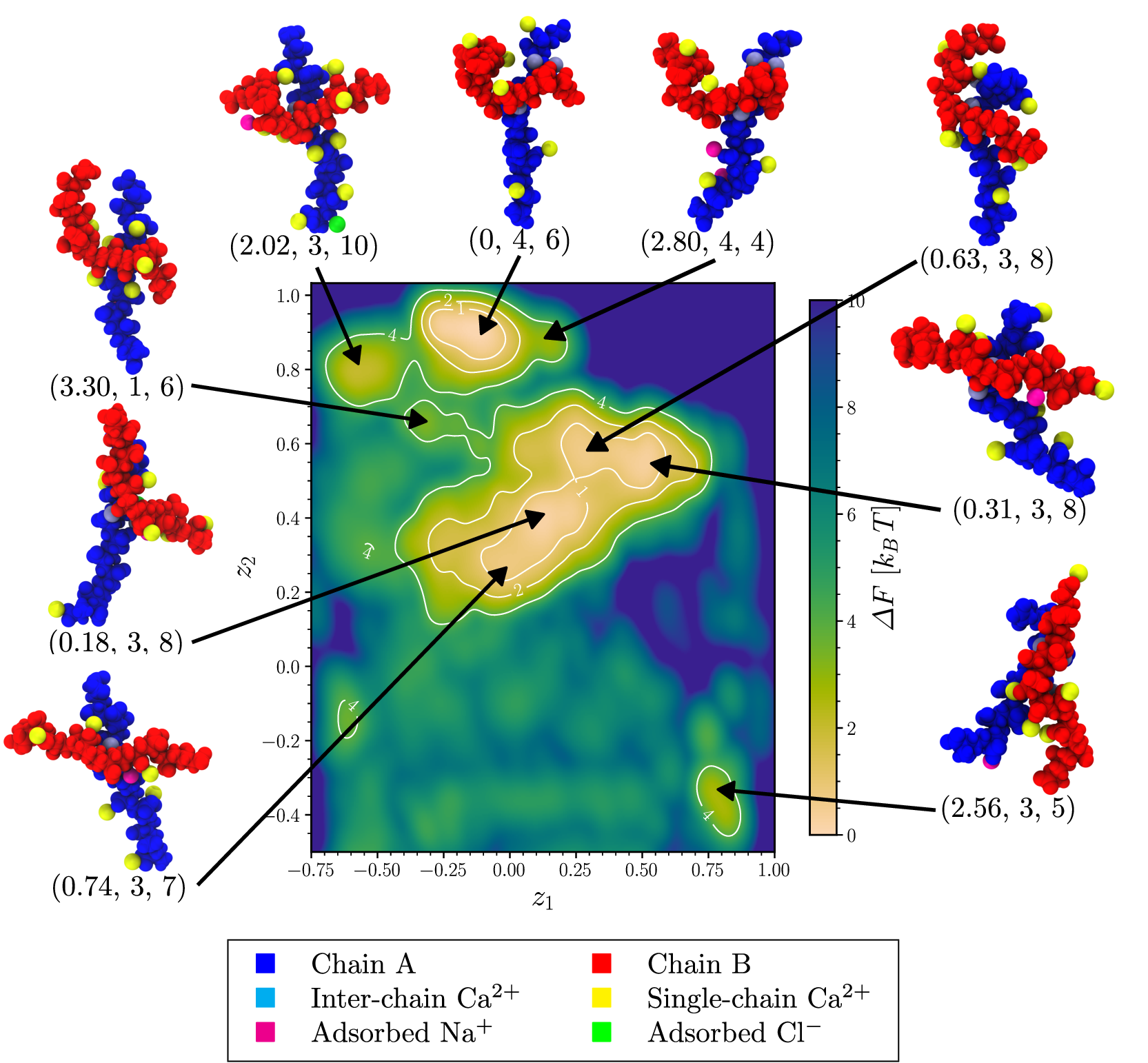}
    \caption{
        Free energy landscape of the autoencoder latent space for two 16-mer PAA chains with 32 Ca$^{2+}$ ions within 1 $k_\mathrm{B} \, T$ of the minimum in the interchain potential of mean force.
        Chain conformations near relevant minima are visualized, along with ions within 0.35 nm of the chains.
        The tuple below each conformation displays the corresponding free energy, the number of bridging calcium ions, and the number of single-chain adsorbed calcium ions.
        Isolines are drawn at 1, 2, and 4 $k_\mathrm{B} \, T$.
    }
    \label{fig:paa_32Ca_ml_conformations}
\end{figure*}
We sought to identify the dominant polymer conformations of the most stable associated states in the 32 Ca$^{2+}$ system that contribute to the two-chain attractive interactions and to explore the relative importance of bridging and single-chain adsorbed Ca$^{2+}$ ions.
Traditional collective variables, such as the radius of gyration or the angle between principal radius of gyration vectors, proved insufficient to discern the dominant conformations due to the multitude of metastable states and the broad distributions of these variables.
To address these challenges, we employed machine learning techniques, specifically an autoencoder, to learn a low-dimensional representation of the conformational space.

Figure~\ref{fig:paa_32Ca_ml_conformations} illustrates the two-dimensional latent space projection ($z_1, z_2$) of the input data, which is reweighted by the Boltzmann factor of the metadynamics bias potential and subsequently transformed into a free energy surface.
The reference state for the free energy surface was set as the most stable minima within the latent space and conformations near relevant minima are rendered.
The chains are colored to indicate the conformations more clearly, and ions within 0.35 nm of the chains are shown.
Water molecules are omitted for clarity.
Remarkably, the autoencoder identifies different ion bridging environments without explicit training on the number of bridging ions or their coordinates.

The most stable conformations reside in the upper energy basin and contain 4 bridging Ca$^{2+}$ with 6 single-chain adsorbed Ca$^{2+}$.
Within this arrangement, one of the chains adopts a collapsed conformation (colored red) and is partially wrapped around the other chain, which is in an extended conformation (colored blue).
The collapsed chain facilitates the formation of ion bridges with the extended chain, and each chain adsorbs three additional Ca$^{2+}$ ions, which serve to neutralize the polymer charge and mitigate unfavorable carboxylate--carboxylate interactions.

The central energy basin hosts a more diverse set of conformations, stabilized by 3 bridging Ca$^{2+}$ ions.
Chains in this region can assume either extended or partially collapsed conformations. The most stable conformation maximizes the number of ion bridges, but interestingly, the autoencoder also identifies conformations with fewer ion bridges that are stabilized by extra single-chain adsorbed Ca$^{2+}$ ions.
Both metastable basins at 0.18 and 0.31 $k_\mathrm{B} \, T$ contain 8 single-chain adsorbed Ca$^{2+}$.
Additional frames depict local minima along the transition path between metastable conformations, as well as a possible transition between the basins of 3 and 4 bridging calcium ions.

The basin with 3 bridging calcium ions at 0.74 $k_\mathrm{B} \, T$ contains 7 single-chain adsorbed Ca$^{2+}$ and highlights the relative importance of ion adsorption and bridging at moderate $N_{\mathrm{Ca}^{2+}}$.
Compared to the minimum free energy conformation observed, the total number of Ca$^{2+}$ on the polyelectrolyte complex is the same in both conformations, but a bridging calcium ion has been replaced by a single-chain adsorbed calcium ion.
The 0.74 $k_\mathrm{B} \, T$ basin also has a relatively extended chain conformations, which likely yields an increase in the chain conformational entropy.
The loss of the ion bridge has a modest impact on free energy ($< 1 \, k_\mathrm{B} \, T$), indicating that the ion bridging does not dominate the free energy landscape.
The cost of losing an ion bridge can be further reduced to 0.18 $k_\mathrm{B} \, T$ by the addition of 1 single-chain adsorbed calcium ion.

\begin{figure}[htpb]
    \centering
    \begin{subfigure}{0.40\linewidth}
        \centering
        \includegraphics[width=\linewidth]{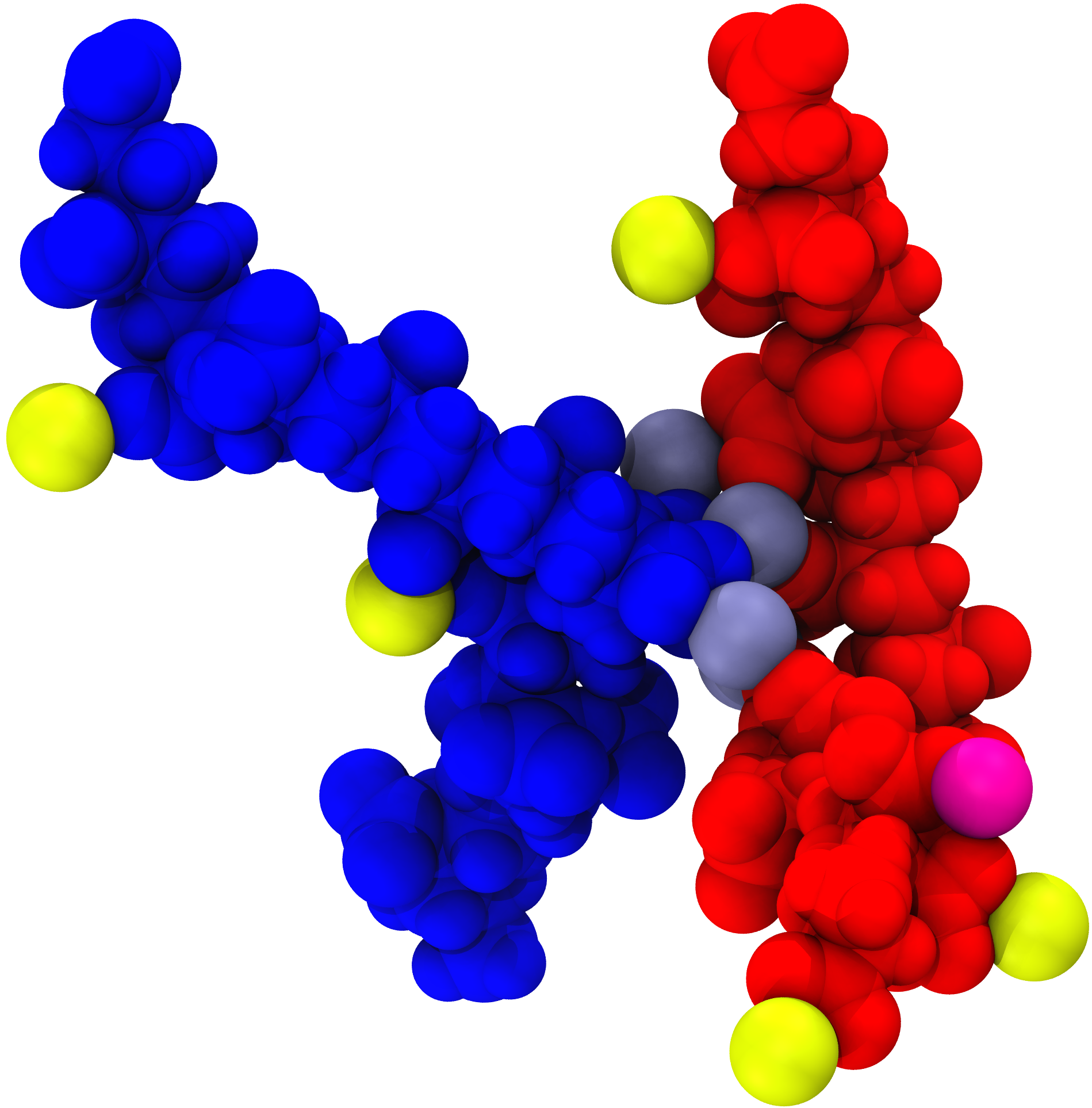}
        \label{sfig:paa_homopolymer_128Ca_conformation_bridging}
    \end{subfigure}
    \hspace*{0.5em}
    \unskip\ \vrule
    \hspace*{0.5em}
    \begin{subfigure}{0.40\linewidth}
        \centering
        \includegraphics[width=\linewidth]{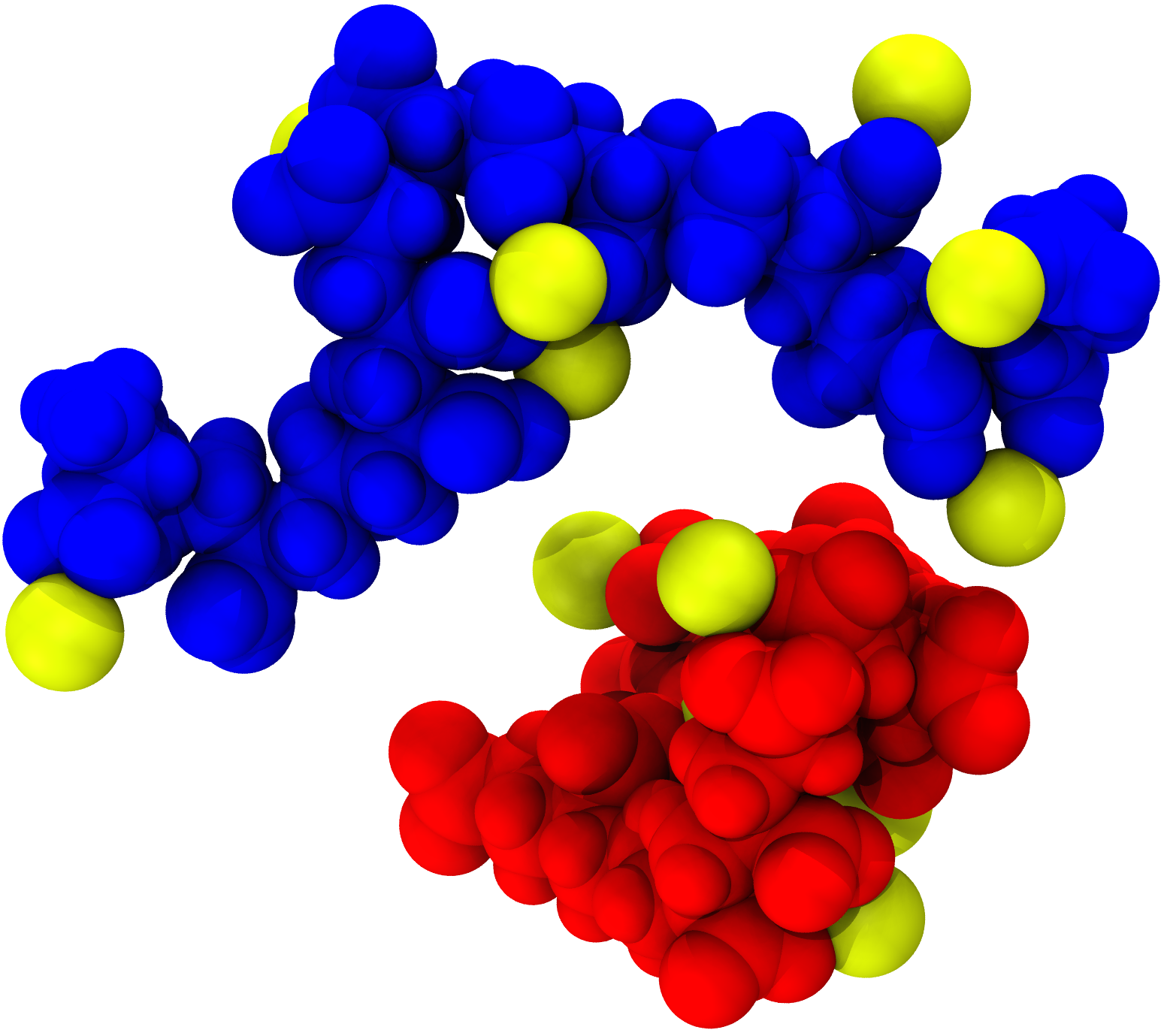}
        \label{sfig:paa_homopolymer_128Ca_conformation_single}
    \end{subfigure}
    \caption{
        Representative conformations of two 16-mer PAA chains with 128 Ca$^{2+}$ ions (within 0.1 $k_\mathrm{B} \, T$ of the minimum in the interchain potential of mean force).
        The visualization depicts the two PAA chains as dark blue and red, Na$^+$ ions as magenta, Ca$^{2+}$ ions as yellow, and interchain bridging Ca$^{2+}$ ions as light blue.
        \textbf{Left panel:} The chains are in direct contact, bridged by 4 Ca$^{2+}$ ions, and have 5 single-chain adsorbed Ca$^{2+}$ ions.
        \textbf{Right panel:} The chains have no bridging Ca$^{2+}$ ions but feature 12 single-chain adsorbed Ca$^{2+}$ ions.
        Note that the red chain's principal axis is aligned into the page for the right panel.
    }
    \label{fig:paa_homopolymer_128Ca_conformations}
\end{figure}
As seen in Figure~\ref{fig:paa_homopolymer_nca_bound}, ion bridges are not dominant at $N_{\mathrm{Ca}^{2+}} = 128$.
However, the presence of ion bridges does not vanish, as seen in Figure~\ref{fig:paa_homopolymer_128Ca_conformations}.
The left panel depicts a conformation with 4 ion bridges and the right panel depicts a conformation with no ion bridges that largely interacts through via solvent mediation.
Both conformations are within 0.1 $k_\mathrm{B} \, T$ of the minimum in the interchain potential of mean force and are highly populated.
However, the majority of conformations in the basin contain no ion bridges, as the increased electrostatic screening reduces the cost of separating the chains and leads to a slightly longer-ranged interaction well (Figure~\ref{fig:paa_homopolymer_interchain_pmf}).
We observe that the increased flexibility of polymer conformations with more Ca$^{2+}$ ions allows for a broader range of conformations, where ion bridges can be destroyed and additional ions adsorbed with almost no free energy difference.
The conformational flexibility may play a role in the salting-in behavior, as the chains can relax into conformations that are more favorable for solvation.

\section{Conclusions}\label{sec:conclusions}

This investigation has uncovered the fundamental factors underpinning the like-charge attraction observed within polyanion solutions mediated by multi-valent cations.
We have found that two primary mechanisms contribute to this phenomenon.

The initial driving force is the increase in the number of favorable polyelectrolyte--ion contacts in chain-associated states when compared to their respective dissociated states (Figure~\ref{fig:paa_homopolymer_2D_fes}).
At lower Ca$^{2+}$ numbers, these contacts lead to ion bridges between the chains, which are important in setting the length-scale of the attraction.
This is manifested in the position of the PMF minima in Figure~\ref{fig:paa_homopolymer_interchain_pmf}.

Our autoencoder analysis revealed that once a sufficient number of ion bridges are formed (3 for the 32 Ca$^{2+}$ system), the free energy difference between forming and breaking additional ion bridges is small compared to thermal energy.
Notably, the ion bridge stability is contingent on electrostatic screening, with higher Ca$^{2+}$ numbers rendering these bridges unnecessary for attraction.
At high ionic strengths, chains are saturated with Ca$^{2+}$ ions and increases in electrostatic screening decrease the favorability of interchain ion bridges, yet the chains still experience a net attraction.
This attraction remains similar in magnitude even when the relative increase of PE--ion contacts in chain-associated states is less pronounced.

The observed attraction at high $\mathrm{Ca}^{2+}$ numbers hints at a second factor driving the attraction: favorable chain--chain interactions that are enabled by the PE--ion contacts on individual chains.
We observed that as the two chains approach one another, they increase the number of PE--ion contacts to stabilize the associated chain states.

Our study is consistent with earlier theoretical findings \cite{de1995precipitation,huang2002polyelectrolytes,liu2003polyelectrolyte,kundagrami2008theory}, which propose that divalent ions can induce chain association.
However, we propose that precipitation might not be dominated by ion bridging induced chain collapse or chain neutralization, as generic polyelectrolyte models may suggest.
As the number of Ca$^{2+}$ increases, chains are saturated with ions, but the saturated PE--ion complex still carries a net-negative charge (Figure~\ref{fig:paa_homopolymer_b2_vs_nca}).
At high Ca$^{2+}$ numbers, electrostatic screening decreases the favorability of direct ion bridging, yet the chains still experience a net attraction due to solvent-mediated interactions between the chains and chelated ions.

Our investigation into chain association builds on the previous research of the Parrinello group \citep{bulo_site_2007,tribello_binding_2009}, focusing on longer chains that can accommodate a wider range of Ca$^{2+}$ adsorption environments, such as ions chelated by multiple, non-neighboring monomers.
Figure~\ref{fig:paa_32Ca_ml_conformations} illustrates the relative favorability of ion bridging, and our observations show that the most stable associated state arises when at least one of the chains adopts an extended conformation, with ion bridges forming exclusively between chains.
Even minor changes in the relative polymer conformations can significantly alter the ion binding environments and the favorability of chain association, as shown in the upper basin of Figure.~\ref{fig:paa_32Ca_ml_conformations}.

Looking forward, potential strategies for maintaining aqueous polyelectrolyte solution stability could involve the introduction of bulky or non-polar side-chains at lower ionic strengths to curtail the short-ranged ion bridging.
At higher ionic strengths, solution stability might be achieved by decreasing the pH to protonate carboxylate groups and subsequently the number of adsorbed ions.
The competition between interchain ion bridging and single-chain ion chelation elucidates a more comprehensive understanding of like-charge attractions in polyanion solutions.


\begin{acknowledgement}
    This work was supported by the Dow Chemical Company through a University Partnership Initiative grant.
    We benefited greatly from the discussions with our Dow collaborators: Thomas Kalantar, Christopher Tucker, Larisa Reyes, and Meng Jing.
    We thank Prof.\ Yasemin Basdogan, Pierre J.\ Walker, and Kayla H.\ Panora for helpful discussions.
\end{acknowledgement}

\endgroup

\bibliography{library.bib, software.bib, other.bib}

\providecommand{\latin}[1]{#1}
\makeatletter
\providecommand{\doi}
  {\begingroup\let\do\@makeother\dospecials
  \catcode`\{=1 \catcode`\}=2 \doi@aux}
\providecommand{\doi@aux}[1]{\endgroup\texttt{#1}}
\makeatother
\providecommand*\mcitethebibliography{\thebibliography}
\csname @ifundefined\endcsname{endmcitethebibliography}  {\let\endmcitethebibliography\endthebibliography}{}
\begin{mcitethebibliography}{77}
\providecommand*\natexlab[1]{#1}
\providecommand*\mciteSetBstSublistMode[1]{}
\providecommand*\mciteSetBstMaxWidthForm[2]{}
\providecommand*\mciteBstWouldAddEndPuncttrue
  {\def\EndOfBibitem{\unskip.}}
\providecommand*\mciteBstWouldAddEndPunctfalse
  {\let\EndOfBibitem\relax}
\providecommand*\mciteSetBstMidEndSepPunct[3]{}
\providecommand*\mciteSetBstSublistLabelBeginEnd[3]{}
\providecommand*\EndOfBibitem{}
\mciteSetBstSublistMode{f}
\mciteSetBstMaxWidthForm{subitem}{(\alph{mcitesubitemcount})}
\mciteSetBstSublistLabelBeginEnd
  {\mcitemaxwidthsubitemform\space}
  {\relax}
  {\relax}

\bibitem[Bolto and Gregory(2007)Bolto, and Gregory]{bolto2007organic}
Bolto,~B.; Gregory,~J. Organic polyelectrolytes in water treatment. \emph{Water Res.} \textbf{2007}, \emph{41}, 2301--2324\relax
\mciteBstWouldAddEndPuncttrue
\mciteSetBstMidEndSepPunct{\mcitedefaultmidpunct}
{\mcitedefaultendpunct}{\mcitedefaultseppunct}\relax
\EndOfBibitem
\bibitem[Lankalapalli and Kolapalli(2009)Lankalapalli, and Kolapalli]{lankalapalli2009polyelectrolyte}
Lankalapalli,~S.; Kolapalli,~V. Polyelectrolyte complexes: A review of their applicability in drug delivery technology. \emph{Indian J. Pharm. Sci.} \textbf{2009}, \emph{71}, 481\relax
\mciteBstWouldAddEndPuncttrue
\mciteSetBstMidEndSepPunct{\mcitedefaultmidpunct}
{\mcitedefaultendpunct}{\mcitedefaultseppunct}\relax
\EndOfBibitem
\bibitem[Reddy and Hoch(2001)Reddy, and Hoch]{reddy_calcite_2001}
Reddy,~M.~M.; Hoch,~A.~R. Calcite crystal growth rate inhibition by polycarboxylic acids. \emph{J. Colloid Interface Sci.} \textbf{2001}, \emph{235}, 365--370, DOI: \doi{10.1006/jcis.2000.7378}, Publisher: Academic Press Inc.\relax
\mciteBstWouldAddEndPunctfalse
\mciteSetBstMidEndSepPunct{\mcitedefaultmidpunct}
{}{\mcitedefaultseppunct}\relax
\EndOfBibitem
\bibitem[Arakawa and Timasheff(1984)Arakawa, and Timasheff]{arakawa1984mechanism}
Arakawa,~T.; Timasheff,~S.~N. Mechanism of protein salting in and salting out by divalent cation salts: balance between hydration and salt binding. \emph{Biochemistry} \textbf{1984}, \emph{23}, 5912--5923\relax
\mciteBstWouldAddEndPuncttrue
\mciteSetBstMidEndSepPunct{\mcitedefaultmidpunct}
{\mcitedefaultendpunct}{\mcitedefaultseppunct}\relax
\EndOfBibitem
\bibitem[Allahyarov \latin{et~al.}(1998)Allahyarov, D'Amico, and Löwen]{allahyarov_attraction_1998}
Allahyarov,~E.; D'Amico,~I.; Löwen,~H. Attraction between {Like}-{Charged} {Macroions} by {Coulomb} {Depletion}. \emph{Phys. Rev. Lett.} \textbf{1998}, \emph{81}, 1334--1337, DOI: \doi{10.1103/PhysRevLett.81.1334}\relax
\mciteBstWouldAddEndPuncttrue
\mciteSetBstMidEndSepPunct{\mcitedefaultmidpunct}
{\mcitedefaultendpunct}{\mcitedefaultseppunct}\relax
\EndOfBibitem
\bibitem[Curtis \latin{et~al.}(2002)Curtis, Ulrich, Montaser, Prausnitz, and Blanch]{curtis_protein-protein_2002}
Curtis,~R.~A.; Ulrich,~J.; Montaser,~A.; Prausnitz,~J.~M.; Blanch,~H.~W. Protein-protein interactions in concentrated electrolyte solutions. \emph{Biotechnol. Bioeng.} \textbf{2002}, \emph{79}, 367--380, DOI: \doi{10.1002/bit.10342}\relax
\mciteBstWouldAddEndPuncttrue
\mciteSetBstMidEndSepPunct{\mcitedefaultmidpunct}
{\mcitedefaultendpunct}{\mcitedefaultseppunct}\relax
\EndOfBibitem
\bibitem[Saluja \latin{et~al.}(2009)Saluja, Crampton, Kras, Fesinmeyer, Remmele, Narhi, Brems, and Gokarn]{saluja_anion_2009}
Saluja,~A.; Crampton,~S.; Kras,~E.; Fesinmeyer,~R.~M.; Remmele,~R.~L.; Narhi,~L.~O.; Brems,~D.~N.; Gokarn,~Y.~R. Anion {Binding} {Mediated} {Precipitation} of a {Peptibody}. \emph{Pharm. Res.} \textbf{2009}, \emph{26}, 152, DOI: \doi{10.1007/s11095-008-9722-0}\relax
\mciteBstWouldAddEndPuncttrue
\mciteSetBstMidEndSepPunct{\mcitedefaultmidpunct}
{\mcitedefaultendpunct}{\mcitedefaultseppunct}\relax
\EndOfBibitem
\bibitem[Li \latin{et~al.}(2017)Li, Girard, Shen, Millan, and Olvera De La~Cruz]{li_strong_2017}
Li,~Y.; Girard,~M.; Shen,~M.; Millan,~J.~A.; Olvera De La~Cruz,~M. Strong attractions and repulsions mediated by monovalent salts. \emph{Proc. Natl. Acad. Sci. U.S.A.} \textbf{2017}, \emph{114}, 11838--11843, DOI: \doi{10.1073/pnas.1713168114}\relax
\mciteBstWouldAddEndPuncttrue
\mciteSetBstMidEndSepPunct{\mcitedefaultmidpunct}
{\mcitedefaultendpunct}{\mcitedefaultseppunct}\relax
\EndOfBibitem
\bibitem[Sinn \latin{et~al.}(2004)Sinn, Dimova, and Antonietti]{sinn_isothermal_2004}
Sinn,~C.~G.; Dimova,~R.; Antonietti,~M. Isothermal {Titration} {Calorimetry} of the {Polyelectrolyte}/{Water} {Interaction} and {Binding} of {Ca} $^{\textrm{2+}}$ : {Effects} {Determining} the {Quality} of {Polymeric} {Scale} {Inhibitors}. \emph{Macromolecules} \textbf{2004}, \emph{37}, 3444--3450, DOI: \doi{10.1021/ma030550s}\relax
\mciteBstWouldAddEndPuncttrue
\mciteSetBstMidEndSepPunct{\mcitedefaultmidpunct}
{\mcitedefaultendpunct}{\mcitedefaultseppunct}\relax
\EndOfBibitem
\bibitem[Gindele \latin{et~al.}(2022)Gindele, Malaszuk, Peter, and Gebauer]{gindele_binding_2022}
Gindele,~M.~B.; Malaszuk,~K.~K.; Peter,~C.; Gebauer,~D. On the {Binding} {Mechanisms} of {Calcium} {Ions} to {Polycarboxylates}: {Effects} of {Molecular} {Weight}, {Side} {Chain}, and {Backbone} {Chemistry}. \emph{Langmuir} \textbf{2022}, \emph{38}, 14409--14421, DOI: \doi{10.1021/acs.langmuir.2c01662}\relax
\mciteBstWouldAddEndPuncttrue
\mciteSetBstMidEndSepPunct{\mcitedefaultmidpunct}
{\mcitedefaultendpunct}{\mcitedefaultseppunct}\relax
\EndOfBibitem
\bibitem[Yu \latin{et~al.}(2004)Yu, Lei, Cheng, and Zhao]{yu_effects_2004}
Yu,~J.; Lei,~M.; Cheng,~B.; Zhao,~X. Effects of {PAA} additive and temperature on morphology of calcium carbonate particles. \emph{J. Solid State Chem.} \textbf{2004}, \emph{177}, 681--689, DOI: \doi{10.1016/j.jssc.2003.08.017}\relax
\mciteBstWouldAddEndPuncttrue
\mciteSetBstMidEndSepPunct{\mcitedefaultmidpunct}
{\mcitedefaultendpunct}{\mcitedefaultseppunct}\relax
\EndOfBibitem
\bibitem[Jada \latin{et~al.}(2007)Jada, Ait~Akbour, Jacquemet, Suau, and Guerret]{jada_effect_2007}
Jada,~A.; Ait~Akbour,~R.; Jacquemet,~C.; Suau,~J.; Guerret,~O. Effect of sodium polyacrylate molecular weight on the crystallogenesis of calcium carbonate. \emph{J. Cryst. Growth} \textbf{2007}, \emph{306}, 373--382, DOI: \doi{10.1016/j.jcrysgro.2007.05.046}\relax
\mciteBstWouldAddEndPuncttrue
\mciteSetBstMidEndSepPunct{\mcitedefaultmidpunct}
{\mcitedefaultendpunct}{\mcitedefaultseppunct}\relax
\EndOfBibitem
\bibitem[Aschauer \latin{et~al.}(2010)Aschauer, Spagnoli, Bowen, and Parker]{aschauer_growth_2010}
Aschauer,~U.; Spagnoli,~D.; Bowen,~P.; Parker,~S.~C. Growth modification of seeded calcite using carboxylic acids: {Atomistic} simulations. \emph{J. Colloid Interface Sci.} \textbf{2010}, \emph{346}, 226--231, DOI: \doi{10.1016/j.jcis.2010.02.057}\relax
\mciteBstWouldAddEndPuncttrue
\mciteSetBstMidEndSepPunct{\mcitedefaultmidpunct}
{\mcitedefaultendpunct}{\mcitedefaultseppunct}\relax
\EndOfBibitem
\bibitem[Huber(1993)]{huber1993calcium}
Huber,~K. Calcium-induced shrinking of polyacrylate chains in aqueous solution. \emph{J. Phys. Chem.} \textbf{1993}, \emph{97}, 9825--9830\relax
\mciteBstWouldAddEndPuncttrue
\mciteSetBstMidEndSepPunct{\mcitedefaultmidpunct}
{\mcitedefaultendpunct}{\mcitedefaultseppunct}\relax
\EndOfBibitem
\bibitem[Schweins and Huber(2001)Schweins, and Huber]{schweins2001collapse}
Schweins,~R.; Huber,~K. Collapse of sodium polyacrylate chains in calcium salt solutions. \emph{Eur. Phys. J. E} \textbf{2001}, \emph{5}, 117--126\relax
\mciteBstWouldAddEndPuncttrue
\mciteSetBstMidEndSepPunct{\mcitedefaultmidpunct}
{\mcitedefaultendpunct}{\mcitedefaultseppunct}\relax
\EndOfBibitem
\bibitem[De~La~Cruz \latin{et~al.}(1995)De~La~Cruz, Belloni, Delsanti, Dalbiez, Spalla, and Drifford]{de1995precipitation}
De~La~Cruz,~M.~O.; Belloni,~L.; Delsanti,~M.; Dalbiez,~J.; Spalla,~O.; Drifford,~M. Precipitation of highly charged polyelectrolyte solutions in the presence of multivalent salts. \emph{J. Chem. Phys.} \textbf{1995}, \emph{103}, 5781--5791\relax
\mciteBstWouldAddEndPuncttrue
\mciteSetBstMidEndSepPunct{\mcitedefaultmidpunct}
{\mcitedefaultendpunct}{\mcitedefaultseppunct}\relax
\EndOfBibitem
\bibitem[Castelnovo \latin{et~al.}(2000)Castelnovo, Sens, and Joanny]{castelnovo_charge_2000}
Castelnovo,~M.; Sens,~P.; Joanny,~J.-F. Charge distribution on annealed polyelectrolytes. \emph{Eur. Phys. J. E} \textbf{2000}, \emph{1}, 115--125, DOI: \doi{10.1007/PL00014591}\relax
\mciteBstWouldAddEndPuncttrue
\mciteSetBstMidEndSepPunct{\mcitedefaultmidpunct}
{\mcitedefaultendpunct}{\mcitedefaultseppunct}\relax
\EndOfBibitem
\bibitem[Huang and Olvera~de La~Cruz(2002)Huang, and Olvera~de La~Cruz]{huang2002polyelectrolytes}
Huang,~C.-I.; Olvera~de La~Cruz,~M. Polyelectrolytes in multivalent salt solutions: Monomolecular versus multimolecular aggregation. \emph{Macromolecules} \textbf{2002}, \emph{35}, 976--986\relax
\mciteBstWouldAddEndPuncttrue
\mciteSetBstMidEndSepPunct{\mcitedefaultmidpunct}
{\mcitedefaultendpunct}{\mcitedefaultseppunct}\relax
\EndOfBibitem
\bibitem[Liu \latin{et~al.}(2003)Liu, Ghosh, and Muthukumar]{liu2003polyelectrolyte}
Liu,~S.; Ghosh,~K.; Muthukumar,~M. Polyelectrolyte solutions with added salt: A simulation study. \emph{J. Chem. Phys.} \textbf{2003}, \emph{119}, 1813--1823\relax
\mciteBstWouldAddEndPuncttrue
\mciteSetBstMidEndSepPunct{\mcitedefaultmidpunct}
{\mcitedefaultendpunct}{\mcitedefaultseppunct}\relax
\EndOfBibitem
\bibitem[Kundagrami and Muthukumar(2008)Kundagrami, and Muthukumar]{kundagrami2008theory}
Kundagrami,~A.; Muthukumar,~M. Theory of competitive counterion adsorption on flexible polyelectrolytes: Divalent salts. \emph{J. Chem. Phys.} \textbf{2008}, \emph{128}\relax
\mciteBstWouldAddEndPuncttrue
\mciteSetBstMidEndSepPunct{\mcitedefaultmidpunct}
{\mcitedefaultendpunct}{\mcitedefaultseppunct}\relax
\EndOfBibitem
\bibitem[Lee and Muthukumar(2009)Lee, and Muthukumar]{lee2009phase}
Lee,~C.-L.; Muthukumar,~M. Phase behavior of polyelectrolyte solutions with salt. \emph{J. Chem. Phys.} \textbf{2009}, \emph{130}\relax
\mciteBstWouldAddEndPuncttrue
\mciteSetBstMidEndSepPunct{\mcitedefaultmidpunct}
{\mcitedefaultendpunct}{\mcitedefaultseppunct}\relax
\EndOfBibitem
\bibitem[Muthukumar(2017)]{muthukumar_50th_2017}
Muthukumar,~M. 50th {Anniversary} {Perspective}: {A} {Perspective} on {Polyelectrolyte} {Solutions}. \emph{Macromolecules} \textbf{2017}, \emph{50}, 9528--9560, DOI: \doi{10.1021/acs.macromol.7b01929}\relax
\mciteBstWouldAddEndPuncttrue
\mciteSetBstMidEndSepPunct{\mcitedefaultmidpunct}
{\mcitedefaultendpunct}{\mcitedefaultseppunct}\relax
\EndOfBibitem
\bibitem[Ha and Liu(1997)Ha, and Liu]{ha_counterion-mediated_1997}
Ha,~B.-Y.; Liu,~A.~J. Counterion-{Mediated} {Attraction} between {Two} {Like}-{Charged} {Rods}. \emph{Phys. Rev. Lett.} \textbf{1997}, \emph{79}, 1289--1292, DOI: \doi{10.1103/PhysRevLett.79.1289}\relax
\mciteBstWouldAddEndPuncttrue
\mciteSetBstMidEndSepPunct{\mcitedefaultmidpunct}
{\mcitedefaultendpunct}{\mcitedefaultseppunct}\relax
\EndOfBibitem
\bibitem[Arenzon \latin{et~al.}(1999)Arenzon, Stilck, and Levin]{arenzon_simple_1999}
Arenzon,~J.; Stilck,~J.; Levin,~Y. Simple model for attraction between like-charged polyions. \emph{Eur. Phys. J. B} \textbf{1999}, \emph{12}, 79--82, DOI: \doi{10.1007/s100510050980}\relax
\mciteBstWouldAddEndPuncttrue
\mciteSetBstMidEndSepPunct{\mcitedefaultmidpunct}
{\mcitedefaultendpunct}{\mcitedefaultseppunct}\relax
\EndOfBibitem
\bibitem[Wittmer \latin{et~al.}(1995)Wittmer, Johner, and Joanny]{wittmer_precipitation_1995}
Wittmer,~J.; Johner,~A.; Joanny,~J.~F. Precipitation of {Polyelectrolytes} in the {Presence} of {Multivalent} {Salts}. \emph{J. phys., II} \textbf{1995}, \emph{5}, 635--654, DOI: \doi{10.1051/jp2:1995154}\relax
\mciteBstWouldAddEndPuncttrue
\mciteSetBstMidEndSepPunct{\mcitedefaultmidpunct}
{\mcitedefaultendpunct}{\mcitedefaultseppunct}\relax
\EndOfBibitem
\bibitem[Solis and De~La~Cruz(2000)Solis, and De~La~Cruz]{solis2000collapse}
Solis,~F.~J.; De~La~Cruz,~M.~O. Collapse of flexible polyelectrolytes in multivalent salt solutions. \emph{J. Chem. Phys.} \textbf{2000}, \emph{112}, 2030--2035\relax
\mciteBstWouldAddEndPuncttrue
\mciteSetBstMidEndSepPunct{\mcitedefaultmidpunct}
{\mcitedefaultendpunct}{\mcitedefaultseppunct}\relax
\EndOfBibitem
\bibitem[Molnar and Rieger(2005)Molnar, and Rieger]{molnar_like-charge_2005}
Molnar,~F.; Rieger,~J. “{Like}-{Charge} {Attraction}” between {Anionic} {Polyelectrolytes}: {Molecular} {Dynamics} {Simulations}. \emph{Langmuir} \textbf{2005}, \emph{21}, 786--789, DOI: \doi{10.1021/la048057c}\relax
\mciteBstWouldAddEndPuncttrue
\mciteSetBstMidEndSepPunct{\mcitedefaultmidpunct}
{\mcitedefaultendpunct}{\mcitedefaultseppunct}\relax
\EndOfBibitem
\bibitem[Bulo \latin{et~al.}(2007)Bulo, Donadio, Laio, Molnar, Rieger, and Parrinello]{bulo_site_2007}
Bulo,~R.~E.; Donadio,~D.; Laio,~A.; Molnar,~F.; Rieger,~J.; Parrinello,~M. “{Site} {Binding}” of {Ca} $^{\textrm{2+}}$ {Ions} to {Polyacrylates} in {Water}: {A} {Molecular} {Dynamics} {Study} of {Coiling} and {Aggregation}. \emph{Macromolecules} \textbf{2007}, \emph{40}, 3437--3442, DOI: \doi{10.1021/ma062467l}\relax
\mciteBstWouldAddEndPuncttrue
\mciteSetBstMidEndSepPunct{\mcitedefaultmidpunct}
{\mcitedefaultendpunct}{\mcitedefaultseppunct}\relax
\EndOfBibitem
\bibitem[Tribello \latin{et~al.}(2009)Tribello, Liew, and Parrinello]{tribello_binding_2009}
Tribello,~G.~A.; Liew,~C.; Parrinello,~M. Binding of {Calcium} and {Carbonate} to {Polyacrylates}. \emph{J. Phys. Chem. B} \textbf{2009}, \emph{113}, 7081--7085, DOI: \doi{10.1021/jp900283d}\relax
\mciteBstWouldAddEndPuncttrue
\mciteSetBstMidEndSepPunct{\mcitedefaultmidpunct}
{\mcitedefaultendpunct}{\mcitedefaultseppunct}\relax
\EndOfBibitem
\bibitem[Mantha \latin{et~al.}()Mantha, Glisman, Yu, and Wang]{mantha2023adsorption}
Mantha,~S.; Glisman,~A.; Yu,~D.; Wang,~Z.-G. Adsorption isotherm and mechanism of Ca$^{2+}$ binding to polyelectrolyte. \emph{Manuscript in review} \relax
\mciteBstWouldAddEndPunctfalse
\mciteSetBstMidEndSepPunct{\mcitedefaultmidpunct}
{}{\mcitedefaultseppunct}\relax
\EndOfBibitem
\bibitem[Berendsen \latin{et~al.}(1995)Berendsen, van~der Spoel, and van Drunen]{berendsen1995gromacs}
Berendsen,~H.~J.; van~der Spoel,~D.; van Drunen,~R. GROMACS: A message-passing parallel molecular dynamics implementation. \emph{Comput. Phys. Commun.} \textbf{1995}, \emph{91}, 43--56\relax
\mciteBstWouldAddEndPuncttrue
\mciteSetBstMidEndSepPunct{\mcitedefaultmidpunct}
{\mcitedefaultendpunct}{\mcitedefaultseppunct}\relax
\EndOfBibitem
\bibitem[Van Der~Spoel \latin{et~al.}(2005)Van Der~Spoel, Lindahl, Hess, Groenhof, Mark, and Berendsen]{van2005gromacs}
Van Der~Spoel,~D.; Lindahl,~E.; Hess,~B.; Groenhof,~G.; Mark,~A.~E.; Berendsen,~H.~J. GROMACS: fast, flexible, and free. \emph{J. Comput. Chem.} \textbf{2005}, \emph{26}, 1701--1718\relax
\mciteBstWouldAddEndPuncttrue
\mciteSetBstMidEndSepPunct{\mcitedefaultmidpunct}
{\mcitedefaultendpunct}{\mcitedefaultseppunct}\relax
\EndOfBibitem
\bibitem[Abraham \latin{et~al.}(2015)Abraham, Murtola, Schulz, P{\'a}ll, Smith, Hess, and Lindahl]{abraham2015gromacs}
Abraham,~M.~J.; Murtola,~T.; Schulz,~R.; P{\'a}ll,~S.; Smith,~J.~C.; Hess,~B.; Lindahl,~E. GROMACS: High performance molecular simulations through multi-level parallelism from laptops to supercomputers. \emph{SoftwareX} \textbf{2015}, \emph{1}, 19--25\relax
\mciteBstWouldAddEndPuncttrue
\mciteSetBstMidEndSepPunct{\mcitedefaultmidpunct}
{\mcitedefaultendpunct}{\mcitedefaultseppunct}\relax
\EndOfBibitem
\bibitem[Bonomi \latin{et~al.}(2009)Bonomi, Branduardi, Bussi, Camilloni, Provasi, Raiteri, Donadio, Marinelli, Pietrucci, Broglia, \latin{et~al.} others]{bonomi2009plumed}
Bonomi,~M.; Branduardi,~D.; Bussi,~G.; Camilloni,~C.; Provasi,~D.; Raiteri,~P.; Donadio,~D.; Marinelli,~F.; Pietrucci,~F.; Broglia,~R.~A.; others PLUMED: A portable plugin for free-energy calculations with molecular dynamics. \emph{Comput. Phys. Commun.} \textbf{2009}, \emph{180}, 1961--1972\relax
\mciteBstWouldAddEndPuncttrue
\mciteSetBstMidEndSepPunct{\mcitedefaultmidpunct}
{\mcitedefaultendpunct}{\mcitedefaultseppunct}\relax
\EndOfBibitem
\bibitem[Tribello \latin{et~al.}(2014)Tribello, Bonomi, Branduardi, Camilloni, and Bussi]{tribello2014plumed}
Tribello,~G.~A.; Bonomi,~M.; Branduardi,~D.; Camilloni,~C.; Bussi,~G. PLUMED 2: New feathers for an old bird. \emph{Comput. Phys. Commun.} \textbf{2014}, \emph{185}, 604--613\relax
\mciteBstWouldAddEndPuncttrue
\mciteSetBstMidEndSepPunct{\mcitedefaultmidpunct}
{\mcitedefaultendpunct}{\mcitedefaultseppunct}\relax
\EndOfBibitem
\bibitem[Bonomi \latin{et~al.}(2019)Bonomi, Bussi, Camilloni, Tribello, Ban{\'a}{\v{s}}, Barducci, Bernetti, Bolhuis, Bottaro, Branduardi, \latin{et~al.} others]{bonomi2019promoting}
Bonomi,~M.; Bussi,~G.; Camilloni,~C.; Tribello,~G.~A.; Ban{\'a}{\v{s}},~P.; Barducci,~A.; Bernetti,~M.; Bolhuis,~P.~G.; Bottaro,~S.; Branduardi,~D.; others Promoting transparency and reproducibility in enhanced molecular simulations. \emph{Nat. Methods} \textbf{2019}, \emph{16}, 670--673\relax
\mciteBstWouldAddEndPuncttrue
\mciteSetBstMidEndSepPunct{\mcitedefaultmidpunct}
{\mcitedefaultendpunct}{\mcitedefaultseppunct}\relax
\EndOfBibitem
\bibitem[Jo \latin{et~al.}(2008)Jo, Kim, Iyer, and Im]{jo2008charmm}
Jo,~S.; Kim,~T.; Iyer,~V.~G.; Im,~W. CHARMM-GUI: a web-based graphical user interface for CHARMM. \emph{J. Comput. Chem.} \textbf{2008}, \emph{29}, 1859--1865\relax
\mciteBstWouldAddEndPuncttrue
\mciteSetBstMidEndSepPunct{\mcitedefaultmidpunct}
{\mcitedefaultendpunct}{\mcitedefaultseppunct}\relax
\EndOfBibitem
\bibitem[Choi \latin{et~al.}(2021)Choi, Park, Park, Kim, Kern, Lee, and Im]{choi2021charmm}
Choi,~Y.~K.; Park,~S.-J.; Park,~S.; Kim,~S.; Kern,~N.~R.; Lee,~J.; Im,~W. CHARMM-GUI polymer builder for modeling and simulation of synthetic polymers. \emph{J. Chem. Theory Comput.} \textbf{2021}, \emph{17}, 2431--2443\relax
\mciteBstWouldAddEndPuncttrue
\mciteSetBstMidEndSepPunct{\mcitedefaultmidpunct}
{\mcitedefaultendpunct}{\mcitedefaultseppunct}\relax
\EndOfBibitem
\bibitem[Berendsen \latin{et~al.}(1987)Berendsen, Grigera, and Straatsma]{berendsen1987missing}
Berendsen,~H.~J.; Grigera,~J.~R.; Straatsma,~T.~P. The missing term in effective pair potentials. \emph{Journal of Physical Chemistry} \textbf{1987}, \emph{91}, 6269--6271\relax
\mciteBstWouldAddEndPuncttrue
\mciteSetBstMidEndSepPunct{\mcitedefaultmidpunct}
{\mcitedefaultendpunct}{\mcitedefaultseppunct}\relax
\EndOfBibitem
\bibitem[Mart{\'\i}nez \latin{et~al.}(2009)Mart{\'\i}nez, Andrade, Birgin, and Mart{\'\i}nez]{martinez2009packmol}
Mart{\'\i}nez,~L.; Andrade,~R.; Birgin,~E.~G.; Mart{\'\i}nez,~J.~M. PACKMOL: A package for building initial configurations for molecular dynamics simulations. \emph{J. Comput. Chem.} \textbf{2009}, \emph{30}, 2157--2164\relax
\mciteBstWouldAddEndPuncttrue
\mciteSetBstMidEndSepPunct{\mcitedefaultmidpunct}
{\mcitedefaultendpunct}{\mcitedefaultseppunct}\relax
\EndOfBibitem
\bibitem[Wang \latin{et~al.}(2004)Wang, Wolf, Caldwell, Kollman, and Case]{wang2004development}
Wang,~J.; Wolf,~R.~M.; Caldwell,~J.~W.; Kollman,~P.~A.; Case,~D.~A. Development and testing of a general amber force field. \emph{J. Comput. Chem.} \textbf{2004}, \emph{25}, 1157--1174\relax
\mciteBstWouldAddEndPuncttrue
\mciteSetBstMidEndSepPunct{\mcitedefaultmidpunct}
{\mcitedefaultendpunct}{\mcitedefaultseppunct}\relax
\EndOfBibitem
\bibitem[Wang \latin{et~al.}(2006)Wang, Wang, Kollman, and Case]{wang2006automatic}
Wang,~J.; Wang,~W.; Kollman,~P.~A.; Case,~D.~A. Automatic atom type and bond type perception in molecular mechanical calculations. \emph{J. Mol. Graph. Model.} \textbf{2006}, \emph{25}, 247--260\relax
\mciteBstWouldAddEndPuncttrue
\mciteSetBstMidEndSepPunct{\mcitedefaultmidpunct}
{\mcitedefaultendpunct}{\mcitedefaultseppunct}\relax
\EndOfBibitem
\bibitem[Mintis and Mavrantzas(2019)Mintis, and Mavrantzas]{mintis_effect_2019}
Mintis,~D.~G.; Mavrantzas,~V.~G. Effect of {pH} and {Molecular} {Length} on the {Structure} and {Dynamics} of {Short} {Poly}(acrylic acid) in {Dilute} {Solution}: {Detailed} {Molecular} {Dynamics} {Study}. \emph{J. Phys. Chem. B} \textbf{2019}, \emph{123}, 4204--4219, DOI: \doi{10.1021/acs.jpcb.9b01696}\relax
\mciteBstWouldAddEndPuncttrue
\mciteSetBstMidEndSepPunct{\mcitedefaultmidpunct}
{\mcitedefaultendpunct}{\mcitedefaultseppunct}\relax
\EndOfBibitem
\bibitem[Martinek \latin{et~al.}(2018)Martinek, Dubou{\'e}-Dijon, Timr, Mason, Baxov{\'a}, Fischer, Schmidt, Pluha{\v{r}}ov{\'a}, and Jungwirth]{martinek_calcium_2018}
Martinek,~T.; Dubou{\'e}-Dijon,~E.; Timr,~{\v{S}}.; Mason,~P.~E.; Baxov{\'a},~K.; Fischer,~H.~E.; Schmidt,~B.; Pluha{\v{r}}ov{\'a},~E.; Jungwirth,~P. Calcium ions in aqueous solutions: Accurate force field description aided by ab initio molecular dynamics and neutron scattering. \emph{J. Chem. Phys.} \textbf{2018}, \emph{148}\relax
\mciteBstWouldAddEndPuncttrue
\mciteSetBstMidEndSepPunct{\mcitedefaultmidpunct}
{\mcitedefaultendpunct}{\mcitedefaultseppunct}\relax
\EndOfBibitem
\bibitem[Leontyev and Stuchebrukhov(2009)Leontyev, and Stuchebrukhov]{leontyev_electronic_2009}
Leontyev,~I.~V.; Stuchebrukhov,~A.~A. Electronic continuum model for molecular dynamics simulations. \emph{J. Chem. Phys.} \textbf{2009}, \emph{130}, 085102, DOI: \doi{10.1063/1.3060164}\relax
\mciteBstWouldAddEndPuncttrue
\mciteSetBstMidEndSepPunct{\mcitedefaultmidpunct}
{\mcitedefaultendpunct}{\mcitedefaultseppunct}\relax
\EndOfBibitem
\bibitem[Duboué-Dijon \latin{et~al.}(2020)Duboué-Dijon, Javanainen, Delcroix, Jungwirth, and Martinez-Seara]{duboue-dijon_practical_2020}
Duboué-Dijon,~E.; Javanainen,~M.; Delcroix,~P.; Jungwirth,~P.; Martinez-Seara,~H. A practical guide to biologically relevant molecular simulations with charge scaling for electronic polarization. \emph{J. Chem. Phys.} \textbf{2020}, \emph{153}, 050901, DOI: \doi{10.1063/5.0017775}\relax
\mciteBstWouldAddEndPuncttrue
\mciteSetBstMidEndSepPunct{\mcitedefaultmidpunct}
{\mcitedefaultendpunct}{\mcitedefaultseppunct}\relax
\EndOfBibitem
\bibitem[Kohagen \latin{et~al.}(2014)Kohagen, Mason, and Jungwirth]{kohagen_accurate_2014}
Kohagen,~M.; Mason,~P.~E.; Jungwirth,~P. Accurate {Description} of {Calcium} {Solvation} in {Concentrated} {Aqueous} {Solutions}. \emph{J. Phys. Chem. B} \textbf{2014}, \emph{118}, 7902--7909, DOI: \doi{10.1021/jp5005693}\relax
\mciteBstWouldAddEndPuncttrue
\mciteSetBstMidEndSepPunct{\mcitedefaultmidpunct}
{\mcitedefaultendpunct}{\mcitedefaultseppunct}\relax
\EndOfBibitem
\bibitem[Dubou{\'e}-Dijon \latin{et~al.}(2018)Dubou{\'e}-Dijon, Delcroix, Martinez-Seara, Hlad{\'\i}lkov{\'a}, Coufal, K\v{r}{\'\i}\v{z}ek, and Jungwirth]{duboue-dijon_binding_2018}
Dubou{\'e}-Dijon,~E.; Delcroix,~P.; Martinez-Seara,~H.; Hlad{\'\i}lkov{\'a},~J.; Coufal,~P.; K\v{r}{\'\i}\v{z}ek,~T.; Jungwirth,~P. Binding of divalent cations to insulin: capillary electrophoresis and molecular simulations. \emph{J. Phys. Chem. B} \textbf{2018}, \emph{122}, 5640--5648\relax
\mciteBstWouldAddEndPuncttrue
\mciteSetBstMidEndSepPunct{\mcitedefaultmidpunct}
{\mcitedefaultendpunct}{\mcitedefaultseppunct}\relax
\EndOfBibitem
\bibitem[Darden \latin{et~al.}(1993)Darden, York, and Pedersen]{darden1993particle}
Darden,~T.; York,~D.; Pedersen,~L. Particle mesh Ewald: An N log (N) method for Ewald sums in large systems. \emph{J. Chem. Phys.} \textbf{1993}, \emph{98}, 10089--10092\relax
\mciteBstWouldAddEndPuncttrue
\mciteSetBstMidEndSepPunct{\mcitedefaultmidpunct}
{\mcitedefaultendpunct}{\mcitedefaultseppunct}\relax
\EndOfBibitem
\bibitem[Essmann \latin{et~al.}(1995)Essmann, Perera, Berkowitz, Darden, Lee, and Pedersen]{essmann1995smooth}
Essmann,~U.; Perera,~L.; Berkowitz,~M.~L.; Darden,~T.; Lee,~H.; Pedersen,~L.~G. A smooth particle mesh Ewald method. \emph{J. Chem. Phys.} \textbf{1995}, \emph{103}, 8577--8593\relax
\mciteBstWouldAddEndPuncttrue
\mciteSetBstMidEndSepPunct{\mcitedefaultmidpunct}
{\mcitedefaultendpunct}{\mcitedefaultseppunct}\relax
\EndOfBibitem
\bibitem[Hess \latin{et~al.}(1997)Hess, Bekker, Berendsen, and Fraaije]{hess1997lincs}
Hess,~B.; Bekker,~H.; Berendsen,~H.~J.; Fraaije,~J.~G. LINCS: A linear constraint solver for molecular simulations. \emph{J. Comput. Chem.} \textbf{1997}, \emph{18}, 1463--1472\relax
\mciteBstWouldAddEndPuncttrue
\mciteSetBstMidEndSepPunct{\mcitedefaultmidpunct}
{\mcitedefaultendpunct}{\mcitedefaultseppunct}\relax
\EndOfBibitem
\bibitem[Nos{\'e}(1984)]{nose1984unified}
Nos{\'e},~S. A unified formulation of the constant temperature molecular dynamics methods. \emph{J. Chem. Phys.} \textbf{1984}, \emph{81}, 511--519\relax
\mciteBstWouldAddEndPuncttrue
\mciteSetBstMidEndSepPunct{\mcitedefaultmidpunct}
{\mcitedefaultendpunct}{\mcitedefaultseppunct}\relax
\EndOfBibitem
\bibitem[Hoover(1985)]{hoover1985canonical}
Hoover,~W.~G. Canonical dynamics: Equilibrium phase-space distributions. \emph{Phys. Rev. A} \textbf{1985}, \emph{31}, 1695\relax
\mciteBstWouldAddEndPuncttrue
\mciteSetBstMidEndSepPunct{\mcitedefaultmidpunct}
{\mcitedefaultendpunct}{\mcitedefaultseppunct}\relax
\EndOfBibitem
\bibitem[Parrinello and Rahman(1981)Parrinello, and Rahman]{parrinello1981polymorphic}
Parrinello,~M.; Rahman,~A. Polymorphic transitions in single crystals: A new molecular dynamics method. \emph{J. Appl. Phys.} \textbf{1981}, \emph{52}, 7182--7190\relax
\mciteBstWouldAddEndPuncttrue
\mciteSetBstMidEndSepPunct{\mcitedefaultmidpunct}
{\mcitedefaultendpunct}{\mcitedefaultseppunct}\relax
\EndOfBibitem
\bibitem[Laio and Parrinello(2002)Laio, and Parrinello]{laio_escaping_2002}
Laio,~A.; Parrinello,~M. Escaping free-energy minima. \emph{Proc. Natl. Acad. Sci. U.S.A.} \textbf{2002}, \emph{99}\relax
\mciteBstWouldAddEndPuncttrue
\mciteSetBstMidEndSepPunct{\mcitedefaultmidpunct}
{\mcitedefaultendpunct}{\mcitedefaultseppunct}\relax
\EndOfBibitem
\bibitem[Barducci \latin{et~al.}(2008)Barducci, Bussi, and Parrinello]{barducci_well-tempered_2008}
Barducci,~A.; Bussi,~G.; Parrinello,~M. Well-tempered metadynamics: {A} smoothly converging and tunable free-energy method. \emph{Phys. Rev. Lett.} \textbf{2008}, \emph{100}, DOI: \doi{10.1103/PhysRevLett.100.020603}, arXiv: 0803.3861\relax
\mciteBstWouldAddEndPuncttrue
\mciteSetBstMidEndSepPunct{\mcitedefaultmidpunct}
{\mcitedefaultendpunct}{\mcitedefaultseppunct}\relax
\EndOfBibitem
\bibitem[Barducci \latin{et~al.}(2011)Barducci, Bonomi, and Parrinello]{barducci_metadynamics_2011}
Barducci,~A.; Bonomi,~M.; Parrinello,~M. Metadynamics. \emph{WIREs Comput. Mol. Sci.} \textbf{2011}, \emph{1}, 826--843, DOI: \doi{10.1002/wcms.31}\relax
\mciteBstWouldAddEndPuncttrue
\mciteSetBstMidEndSepPunct{\mcitedefaultmidpunct}
{\mcitedefaultendpunct}{\mcitedefaultseppunct}\relax
\EndOfBibitem
\bibitem[Wang \latin{et~al.}(2011)Wang, Friesner, and Berne]{wang_replica_2011}
Wang,~L.; Friesner,~R.~A.; Berne,~B.~J. Replica {Exchange} with {Solute} {Scaling}: {A} {More} {Efficient} {Version} of {Replica} {Exchange} with {Solute} {Tempering} ({REST2}). \emph{J. Phys. Chem. B} \textbf{2011}, \emph{115}, 9431--9438, DOI: \doi{10.1021/jp204407d}\relax
\mciteBstWouldAddEndPuncttrue
\mciteSetBstMidEndSepPunct{\mcitedefaultmidpunct}
{\mcitedefaultendpunct}{\mcitedefaultseppunct}\relax
\EndOfBibitem
\bibitem[Bussi(2014)]{bussi_hamiltonian_2014}
Bussi,~G. Hamiltonian replica exchange in {GROMACS}: a flexible implementation. \emph{Mol. Phys.} \textbf{2014}, \emph{112}, 379--384, DOI: \doi{10.1080/00268976.2013.824126}\relax
\mciteBstWouldAddEndPuncttrue
\mciteSetBstMidEndSepPunct{\mcitedefaultmidpunct}
{\mcitedefaultendpunct}{\mcitedefaultseppunct}\relax
\EndOfBibitem
\bibitem[Park \latin{et~al.}(2012)Park, Zhu, and Yethiraj]{park_atomistic_2012}
Park,~S.; Zhu,~X.; Yethiraj,~A. Atomistic {Simulations} of {Dilute} {Polyelectrolyte} {Solutions}. \emph{J. Phys. Chem. B} \textbf{2012}, \emph{116}, 4319--4327, DOI: \doi{10.1021/jp208138t}\relax
\mciteBstWouldAddEndPuncttrue
\mciteSetBstMidEndSepPunct{\mcitedefaultmidpunct}
{\mcitedefaultendpunct}{\mcitedefaultseppunct}\relax
\EndOfBibitem
\bibitem[Robbins and Wang(2013)Robbins, and Wang]{robbins_effect_2013}
Robbins,~T.~J.; Wang,~Y. Effect of initial ion positions on the interactions of monovalent and divalent ions with a {DNA} duplex as revealed with atomistic molecular dynamics simulations. \emph{J. Biomol. Struct. Dyn.} \textbf{2013}, \emph{31}, 1311--1323, DOI: \doi{10.1080/07391102.2012.732344}\relax
\mciteBstWouldAddEndPuncttrue
\mciteSetBstMidEndSepPunct{\mcitedefaultmidpunct}
{\mcitedefaultendpunct}{\mcitedefaultseppunct}\relax
\EndOfBibitem
\bibitem[Michaud-Agrawal \latin{et~al.}(2011)Michaud-Agrawal, Denning, Woolf, and Beckstein]{michaud2011mdanalysis}
Michaud-Agrawal,~N.; Denning,~E.~J.; Woolf,~T.~B.; Beckstein,~O. MDAnalysis: a toolkit for the analysis of molecular dynamics simulations. \emph{J. Comput. Chem.} \textbf{2011}, \emph{32}, 2319--2327\relax
\mciteBstWouldAddEndPuncttrue
\mciteSetBstMidEndSepPunct{\mcitedefaultmidpunct}
{\mcitedefaultendpunct}{\mcitedefaultseppunct}\relax
\EndOfBibitem
\bibitem[Gowers \latin{et~al.}(2016)Gowers, Linke, Barnoud, Reddy, Melo, Seyler, Domanski, Dotson, Buchoux, Kenney, \latin{et~al.} others]{gowers2016mdanalysis}
Gowers,~R.~J.; Linke,~M.; Barnoud,~J.; Reddy,~T.~J.; Melo,~M.~N.; Seyler,~S.~L.; Domanski,~J.; Dotson,~D.~L.; Buchoux,~S.; Kenney,~I.~M.; others MDAnalysis: a Python package for the rapid analysis of molecular dynamics simulations. Proceedings of the 15th python in science conference. 2016; p 105\relax
\mciteBstWouldAddEndPuncttrue
\mciteSetBstMidEndSepPunct{\mcitedefaultmidpunct}
{\mcitedefaultendpunct}{\mcitedefaultseppunct}\relax
\EndOfBibitem
\bibitem[Humphrey \latin{et~al.}(1996)Humphrey, Dalke, and Schulten]{HUMP96}
Humphrey,~W.; Dalke,~A.; Schulten,~K. {VMD} -- {V}isual {M}olecular {D}ynamics. \emph{J. Mol. Graph.} \textbf{1996}, \emph{14}, 33--38\relax
\mciteBstWouldAddEndPuncttrue
\mciteSetBstMidEndSepPunct{\mcitedefaultmidpunct}
{\mcitedefaultendpunct}{\mcitedefaultseppunct}\relax
\EndOfBibitem
\bibitem[Stone(1998)]{STON1998}
Stone,~J. {\em An Efficient Library for Parallel Ray Tracing and Animation}. M.Sc.\ thesis, Computer Science Department, University of Missouri-Rolla, 1998\relax
\mciteBstWouldAddEndPuncttrue
\mciteSetBstMidEndSepPunct{\mcitedefaultmidpunct}
{\mcitedefaultendpunct}{\mcitedefaultseppunct}\relax
\EndOfBibitem
\bibitem[Neal \latin{et~al.}(1998)Neal, Asthagiri, and Lenhoff]{neal_molecular_1998}
Neal,~B.; Asthagiri,~D.; Lenhoff,~A. Molecular {Origins} of {Osmotic} {Second} {Virial} {Coefficients} of {Proteins}. \emph{Biophys. J.} \textbf{1998}, \emph{75}, 2469--2477, DOI: \doi{10.1016/S0006-3495(98)77691-X}\relax
\mciteBstWouldAddEndPuncttrue
\mciteSetBstMidEndSepPunct{\mcitedefaultmidpunct}
{\mcitedefaultendpunct}{\mcitedefaultseppunct}\relax
\EndOfBibitem
\bibitem[Lund and Jönsson(2003)Lund, and Jönsson]{lund_mesoscopic_2003}
Lund,~M.; Jönsson,~B. A {Mesoscopic} {Model} for {Protein}-{Protein} {Interactions} in {Solution}. \emph{Biophys. J.} \textbf{2003}, \emph{85}, 2940--2947, DOI: \doi{10.1016/S0006-3495(03)74714-6}\relax
\mciteBstWouldAddEndPuncttrue
\mciteSetBstMidEndSepPunct{\mcitedefaultmidpunct}
{\mcitedefaultendpunct}{\mcitedefaultseppunct}\relax
\EndOfBibitem
\bibitem[Stark \latin{et~al.}(2013)Stark, Andrews, and Elcock]{stark_toward_2013}
Stark,~A.~C.; Andrews,~C.~T.; Elcock,~A.~H. Toward {Optimized} {Potential} {Functions} for {Protein}–{Protein} {Interactions} in {Aqueous} {Solutions}: {Osmotic} {Second} {Virial} {Coefficient} {Calculations} {Using} the {MARTINI} {Coarse}-{Grained} {Force} {Field}. \emph{J. Chem. Theory Comput.} \textbf{2013}, \emph{9}, 4176--4185, DOI: \doi{10.1021/ct400008p}\relax
\mciteBstWouldAddEndPuncttrue
\mciteSetBstMidEndSepPunct{\mcitedefaultmidpunct}
{\mcitedefaultendpunct}{\mcitedefaultseppunct}\relax
\EndOfBibitem
\bibitem[Neal \latin{et~al.}(1998)Neal, Asthagiri, and Lenhoff]{neal1998molecular}
Neal,~B.; Asthagiri,~D.; Lenhoff,~A. Molecular origins of osmotic second virial coefficients of proteins. \emph{Biophys. J.} \textbf{1998}, \emph{75}, 2469--2477\relax
\mciteBstWouldAddEndPuncttrue
\mciteSetBstMidEndSepPunct{\mcitedefaultmidpunct}
{\mcitedefaultendpunct}{\mcitedefaultseppunct}\relax
\EndOfBibitem
\bibitem[Hill(2013)]{hill2013statistical}
Hill,~T.~L. \emph{Statistical mechanics: principles and selected applications}; Courier Corporation, 2013\relax
\mciteBstWouldAddEndPuncttrue
\mciteSetBstMidEndSepPunct{\mcitedefaultmidpunct}
{\mcitedefaultendpunct}{\mcitedefaultseppunct}\relax
\EndOfBibitem
\bibitem[Lemke and Peter(2019)Lemke, and Peter]{lemke_encodermap_2019}
Lemke,~T.; Peter,~C. {EncoderMap}: {Dimensionality} {Reduction} and {Generation} of {Molecule} {Conformations}. \emph{J. Chem. Theory Comput.} \textbf{2019}, \emph{15}, 1209--1215, DOI: \doi{10.1021/acs.jctc.8b00975}\relax
\mciteBstWouldAddEndPuncttrue
\mciteSetBstMidEndSepPunct{\mcitedefaultmidpunct}
{\mcitedefaultendpunct}{\mcitedefaultseppunct}\relax
\EndOfBibitem
\bibitem[Lemke \latin{et~al.}(2019)Lemke, Berg, Jain, and Peter]{lemke_encodermapii_2019}
Lemke,~T.; Berg,~A.; Jain,~A.; Peter,~C. {EncoderMap}({II}): {Visualizing} {Important} {Molecular} {Motions} with {Improved} {Generation} of {Protein} {Conformations}. \emph{J Chem. Inf. Model.} \textbf{2019}, \emph{59}, 4550--4560, DOI: \doi{10.1021/acs.jcim.9b00675}\relax
\mciteBstWouldAddEndPuncttrue
\mciteSetBstMidEndSepPunct{\mcitedefaultmidpunct}
{\mcitedefaultendpunct}{\mcitedefaultseppunct}\relax
\EndOfBibitem
\bibitem[Bandyopadhyay and Mondal(2021)Bandyopadhyay, and Mondal]{bandyopadhyay_deep_2021}
Bandyopadhyay,~S.; Mondal,~J. A deep autoencoder framework for discovery of metastable ensembles in biomacromolecules. \emph{J. Chem. Phys.} \textbf{2021}, \emph{155}, DOI: \doi{10.1063/5.0059965}, arXiv: 2106.00724 Publisher: American Institute of Physics Inc.\relax
\mciteBstWouldAddEndPunctfalse
\mciteSetBstMidEndSepPunct{\mcitedefaultmidpunct}
{}{\mcitedefaultseppunct}\relax
\EndOfBibitem
\bibitem[Glielmo \latin{et~al.}(2021)Glielmo, Husic, Rodriguez, Clementi, Noé, and Laio]{glielmo_unsupervised_2021}
Glielmo,~A.; Husic,~B.~E.; Rodriguez,~A.; Clementi,~C.; Noé,~F.; Laio,~A. Unsupervised {Learning} {Methods} for {Molecular} {Simulation} {Data}. \emph{Chem. Rev.} \textbf{2021}, \emph{121}, 9722--9758, DOI: \doi{10.1021/acs.chemrev.0c01195}\relax
\mciteBstWouldAddEndPuncttrue
\mciteSetBstMidEndSepPunct{\mcitedefaultmidpunct}
{\mcitedefaultendpunct}{\mcitedefaultseppunct}\relax
\EndOfBibitem
\bibitem[Paszke \latin{et~al.}(2017)Paszke, Gross, Chintala, Chanan, Yang, DeVito, Lin, Desmaison, Antiga, and Lerer]{paszke2017automatic}
Paszke,~A.; Gross,~S.; Chintala,~S.; Chanan,~G.; Yang,~E.; DeVito,~Z.; Lin,~Z.; Desmaison,~A.; Antiga,~L.; Lerer,~A. Automatic differentiation in pytorch. \textbf{2017}, \relax
\mciteBstWouldAddEndPunctfalse
\mciteSetBstMidEndSepPunct{\mcitedefaultmidpunct}
{}{\mcitedefaultseppunct}\relax
\EndOfBibitem
\bibitem[Paszke \latin{et~al.}(2019)Paszke, Gross, Massa, Lerer, Bradbury, Chanan, Killeen, Lin, Gimelshein, Antiga, \latin{et~al.} others]{paszke2019pytorch}
Paszke,~A.; Gross,~S.; Massa,~F.; Lerer,~A.; Bradbury,~J.; Chanan,~G.; Killeen,~T.; Lin,~Z.; Gimelshein,~N.; Antiga,~L.; others Pytorch: An imperative style, high-performance deep learning library. \emph{Adv. Neural Inf. Process. Syst.} \textbf{2019}, \emph{32}\relax
\mciteBstWouldAddEndPuncttrue
\mciteSetBstMidEndSepPunct{\mcitedefaultmidpunct}
{\mcitedefaultendpunct}{\mcitedefaultseppunct}\relax
\EndOfBibitem
\end{mcitethebibliography}

%
%

\end{document}


\begingroup
\section{Polyelectrolyte--Ion Structure}\label{apsec:rdf}

In this section, we present additional details on the structure of the polyelectrolyte--ion complexes for the 16-mer poly(acrylic acid) (PAA) systems.
The potential of mean force (PMF) curves as a function of the interchain distance shown in the main text are coarse-grained metrics of the interchain interactions and do not provide information about the local structure of the complexes.
In Figure \ref{fig:paa_homopolymer_chain_pairdists}, we present the radial distribution function (RDF) of side-chain carboxylate carbon atoms, denoted C$_\mathrm{cb}$, on one chain (A) with respect to C$_\mathrm{cb}$ atoms on the other chain (B).
\nomenclature[RDF]{RDF}{radial distribution function}
\begin{figure*}[!htb]
    \centering
    \begin{subfigure}{0.55\linewidth}
        \centering
        \includegraphics[width=\linewidth]{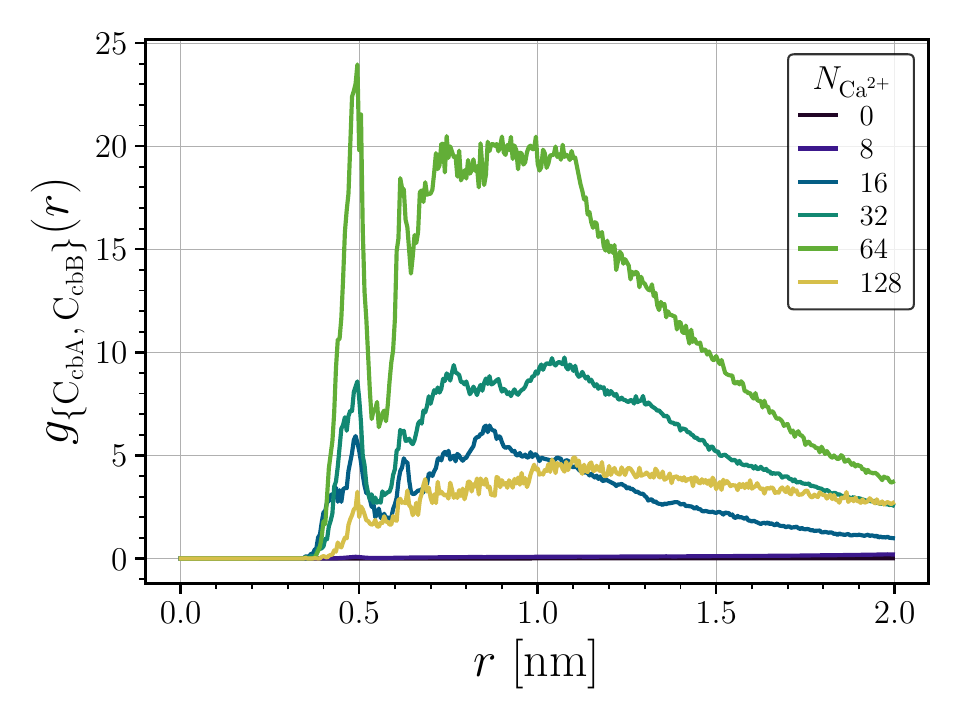}
    \end{subfigure}
    \caption{
        RDF of carboxylate carbon (C$_\mathrm{cb}$) atoms on one chain with respect to carboxylate carbon atoms on the other chain.
    }
    \label{fig:paa_homopolymer_chain_pairdists}
\end{figure*}
As the C$_\mathrm{cb}$ atoms are on the side-chains and directly involved in the ion binding/bridging, they provide a local measure of the interchain interaction distance.
The carboxylate carbon RDFs show a well-defined peak at 0.5 nm that is representative of the distance between carboxylate groups bridged by a single Ca$^{2+}$ ion.
Notably, the height of this peak shows the same non-monotonic trend as the number of bridging Ca$^{2+}$ ions seen in the main text with increasing Ca$^{2+}$ ions.

\begin{figure*}[!htb]
    \centering
    \includegraphics[width=0.495\linewidth]{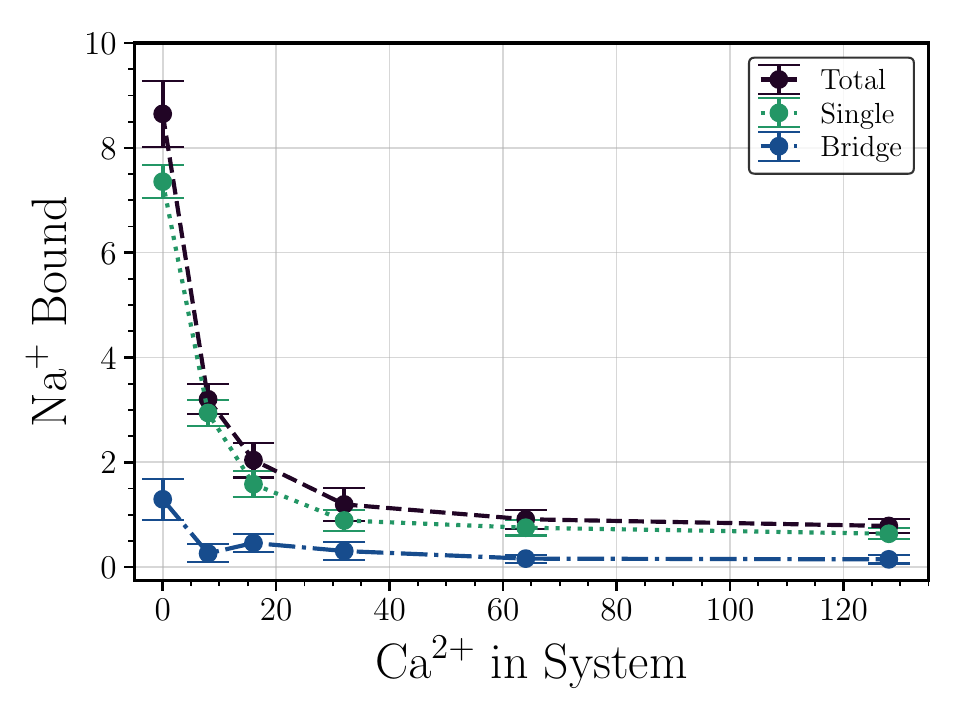}
    \caption{
        Number of sodium ions bound to PAA carboxylate groups as a function of the number of calcium ions in the system.
        Data plotted within 1 $k_\mathrm{B} T$ of the minimum in the interchain potential of mean force.
        Ions bound to a single chain are green, and bridging ions are blue.
        The 0 Ca$^{2+}$ data comes from associated states where the interchain distance was less than 2 nm, as the PMF was repulsive at all distances.
        The dotted lines are guides to the eye.
    }
    \label{fig:paa_homopolymer_na_adsorbed}
\end{figure*}
Ca$^{2+}$ outcompete Na$^+$ for binding to the carboxylate groups, and the number of Na$^+$ bound to the carboxylate groups decreases with increasing Ca$^{2+}$, as shown in Figure \ref{fig:paa_homopolymer_na_adsorbed}.
Even at high Ca$^{2+}$ numbers, one Na$^+$ remains bound to the polyelectrolyte--ion complex on average and does not contribute significantly to the ion bridging behavior.

Further RDFs illustrating the ion binding to carboxylate carbons and oxygens (O$_\mathrm{cb}$) are shown in Figures \ref{fig:paa_homopolymer_ca_pairdists} and \ref{fig:paa_homopolymer_na_pairdists}.
Two dominant peaks are observed about carboxylate carbon atoms corresponding to bidentate and monodentate chelation, respectively.
\begin{figure*}[!htb]
    \centering
    \begin{subfigure}{0.495\linewidth}
        \centering
        \includegraphics[width=\linewidth]{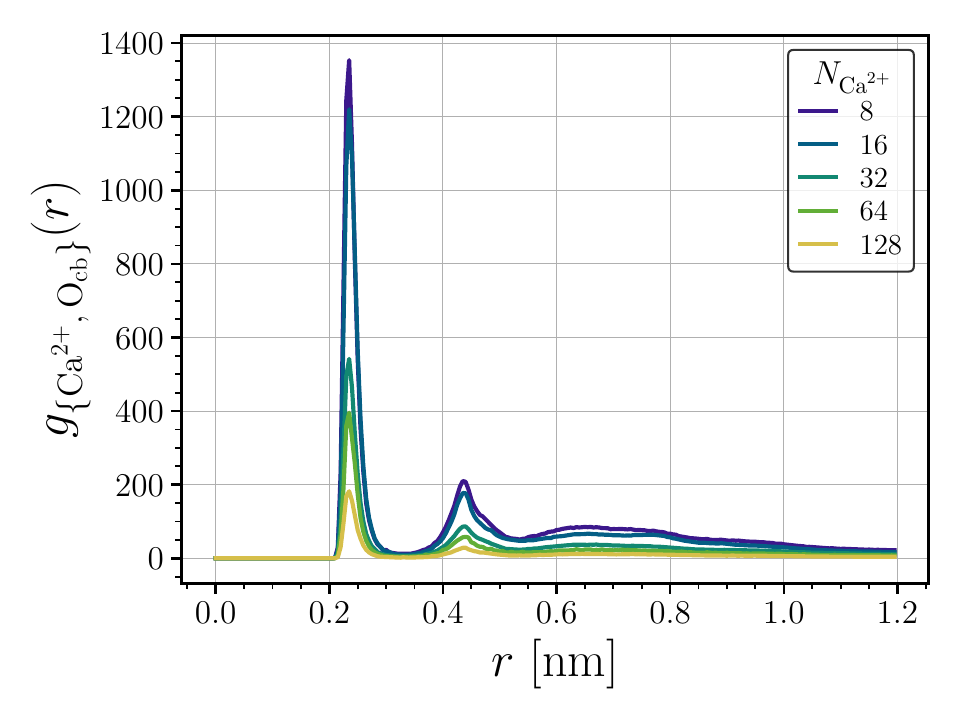}
    \end{subfigure}
    \begin{subfigure}{0.495\linewidth}
        \centering
        \includegraphics[width=\linewidth]{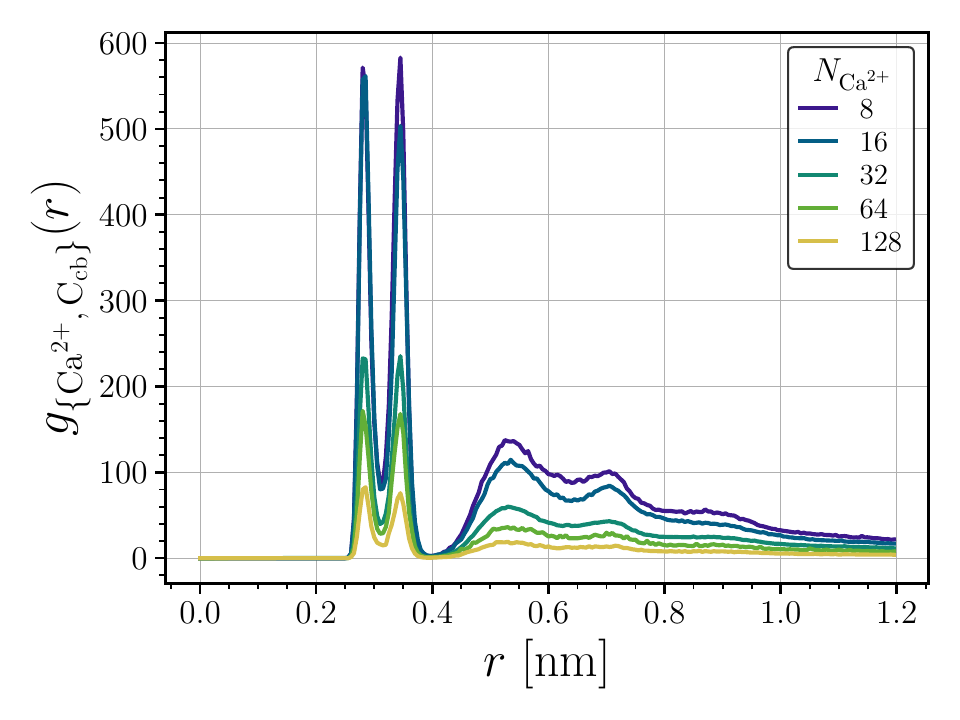}
    \end{subfigure}
    \caption{
        \textbf{Left panel:} RDF of calcium ions around carboxylate oxygen atoms.
        \textbf{Right panel:} RDF of calcium ions around carboxylate carbon atoms.
    }
    \label{fig:paa_homopolymer_ca_pairdists}
\end{figure*}
\begin{figure*}[!htb]
    \centering
    \begin{subfigure}{0.495\linewidth}
        \centering
        \includegraphics[width=\linewidth]{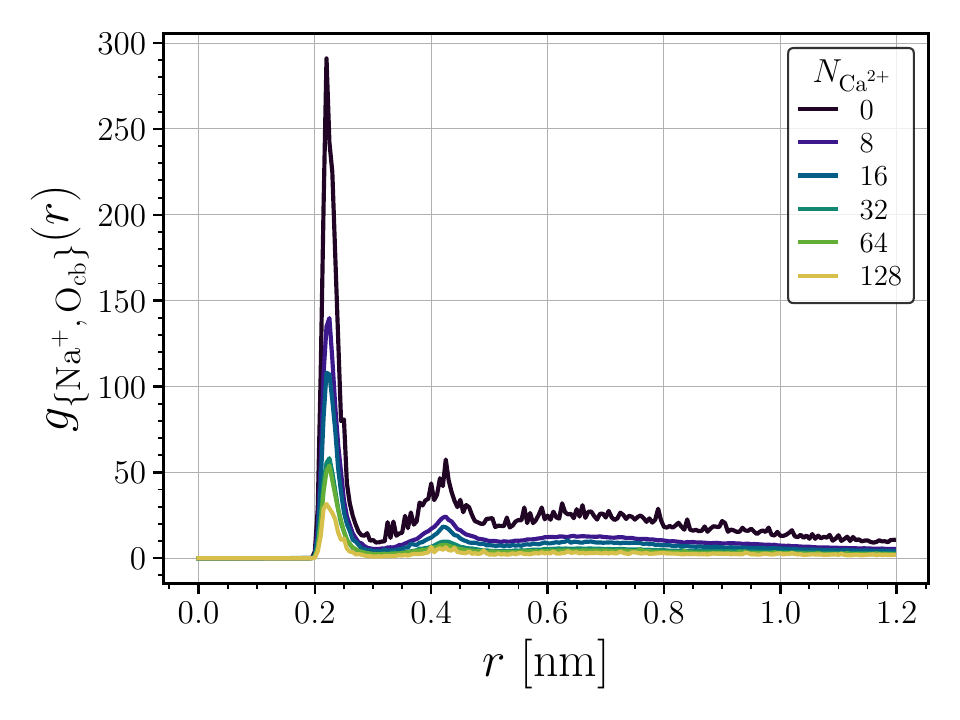}
    \end{subfigure}
    \begin{subfigure}{0.495\linewidth}
        \centering
        \includegraphics[width=\linewidth]{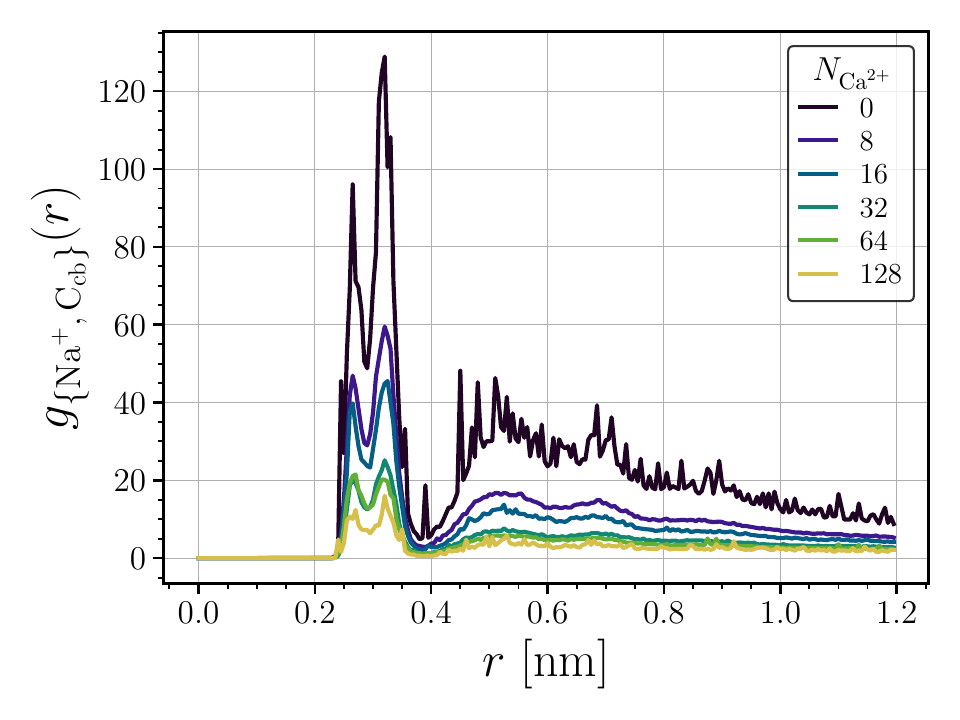}
    \end{subfigure}
    \caption{
        \textbf{Left panel:} RDF of sodium ions around carboxylate oxygen atoms.
        \textbf{Right panel:} RDF of sodium ions around carboxylate carbon atoms.
    }
    \label{fig:paa_homopolymer_na_pairdists}
\end{figure*}

\section{Additional Free Energy Surfaces}\label{apsec:additional_fes}

We show the free energy surfaces for the number of calcium and carboxylate oxygen contacts as a function of the interchain center of mass distance for systems not shown in the main text in Figure \ref{fig:paa_homopolymer_2D_fes}.
The 8 Ca$^{2+}$ system favors dissociated chain states, as there are not enough Ca$^{2+}$ ions to form sufficient numbers of polymer--ion contacts to stabilize the two-chain complex.
The 16 and 64 Ca$^{2+}$ systems exhibit a qualitatively similar free energy surface to the 32 Ca$^{2+}$ system in the main text.
\begin{figure*}[htb]
    \centering
    \begin{subfigure}{0.495\linewidth}
        \centering
        \captionsetup{justification=centering}
        \caption*{\large 8 Ca$^{2+}$}
        \vspace*{-1em}
        \label{sfig:paa_homopolymer_8Ca_2D_fes_distance_cn}
        \includegraphics[width=\linewidth]{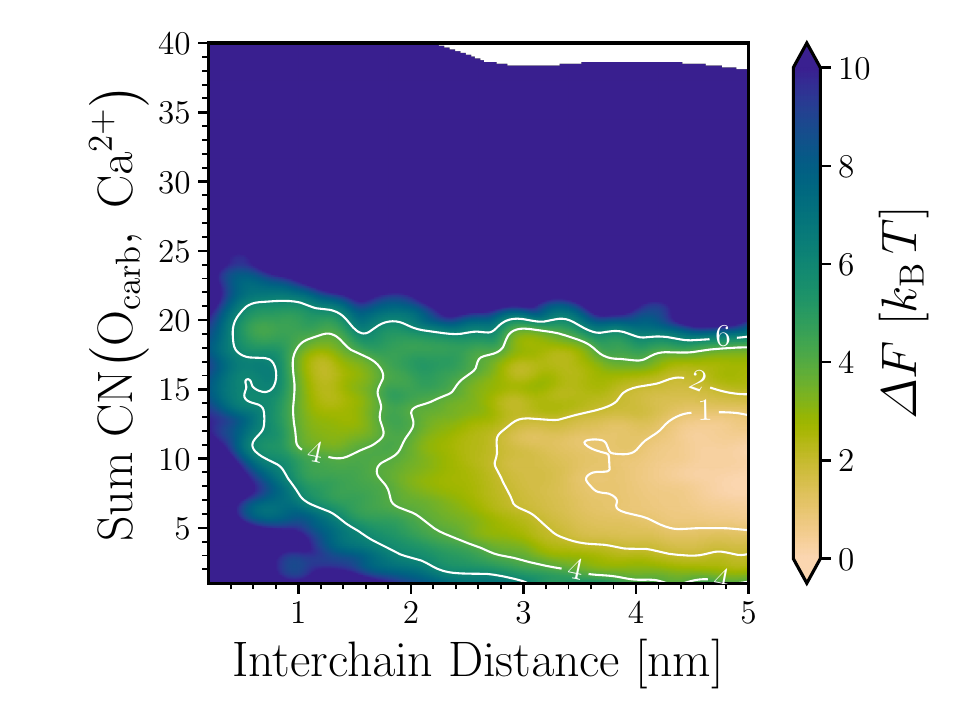}
    \end{subfigure}
    \begin{subfigure}{0.495\linewidth}
        \centering
        \captionsetup{justification=centering}
        \caption*{\large 16 Ca$^{2+}$}
        \vspace*{-1em}
        \label{sfig:paa_homopolymer_16Ca_2D_fes_distance_cn}
        \includegraphics[width=\linewidth]{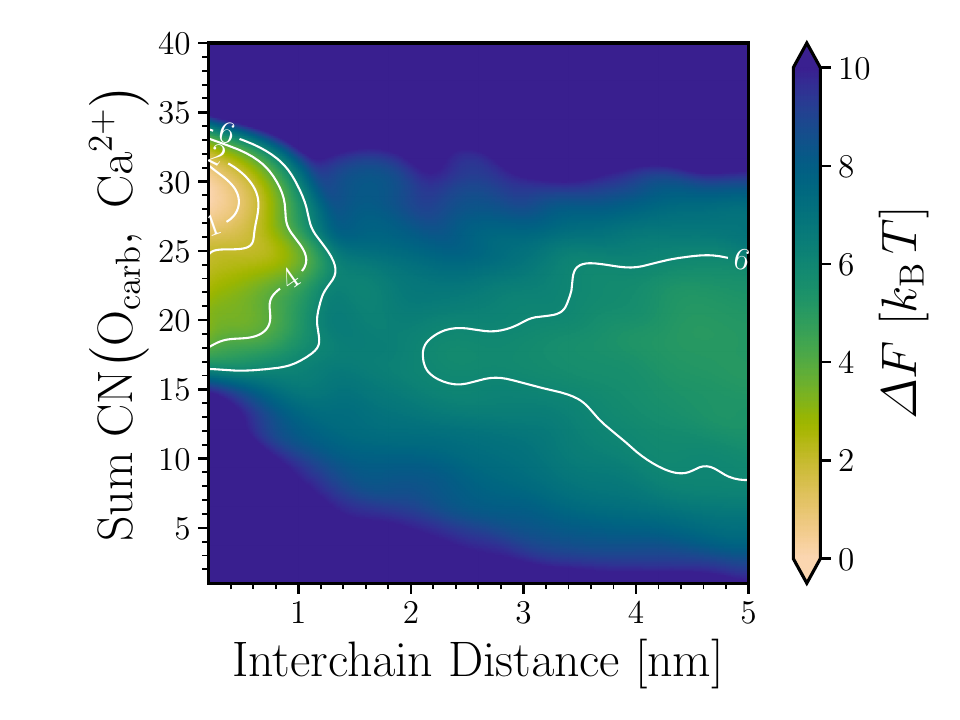}
    \end{subfigure}
    \\
    \begin{subfigure}{0.495\linewidth}
        \centering
        \captionsetup{justification=centering}
        \caption*{\large 64 Ca$^{2+}$}
        \vspace*{-1em}
        \label{sfig:paa_homopolymer_128Ca_2D_fes_distance_cn}
        \includegraphics[width=\linewidth]{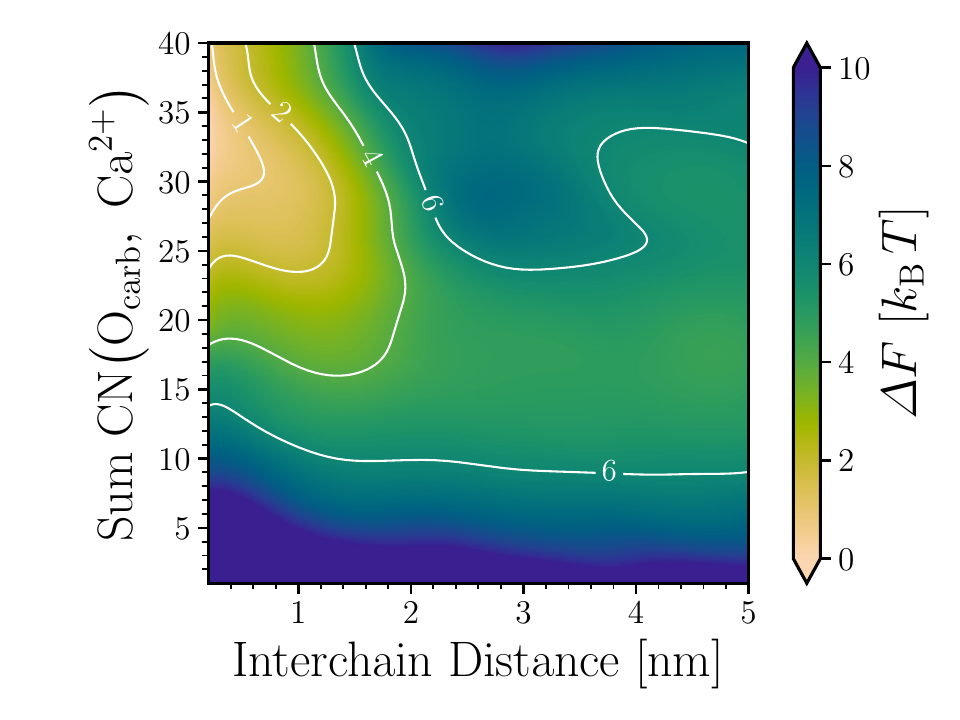}
    \end{subfigure}
    \caption{
        Two-dimensional free energy surfaces for two 16-mer PAA chains with 8 (\textbf{upper left panel}),
        16 (\textbf{upper right panel}), and 64 (\textbf{bottom panel}) Ca$^{2+}$ ions.
        The horizontal axis denotes the chain center of mass distance, and the vertical axis denotes the summed coordination number of carboxylate oxygen atoms around Ca$^{2+}$ ions.
        Isolines are drawn at 1, 2, 4, and 6 $k_\mathrm{B} T$.
    }
    \label{fig:paa_homopolymer_2D_fes}
\end{figure*}
\section{Autoencoder}\label{apsec:autoencoder}

As discussed in the main text, the input data to the autoencoder (AE) consisted of the pairwise distances between all backbone C$_\alpha$ atoms for the 32 Ca$^{2+}$ system.
We further restricted the input data to frames where the interchain distance was within 1 $k_\mathrm{B} T$ of the minimum in the interchain PMF to focus on the dominant conformational states.
The distances were then scaled to the range $[0, 1]$ to improve the training of the AE.
Our autoencoder (AE) architecture consisted of an encoder and a decoder connected by a latent space.
The encoder consisted of 3 fully connected hidden layers of 496, 128, and 32 neurons, respectively, with hyperbolic tangent activation functions.
The decoder was a mirror image of the encoder with the same number of layers and neurons but had an additional output layer with sigmoid activation to scale the output to the range $[0, 1]$.
Weights were initialized with the Xavier uniform initializer.

We determined the optimal latent space by minimizing the reconstruction error of the AE with a fixed latent space dimension of 2 for visualization purposes.
The training dynamics of the AE, represented by the loss function during the training process, are shown in Figure \ref{fig:loss-dynamics}.
The loss function converges during training, indicating successful learning and feature extraction.
\begin{figure}[!htb]
    \centering
    \includegraphics[width=0.5\textwidth]{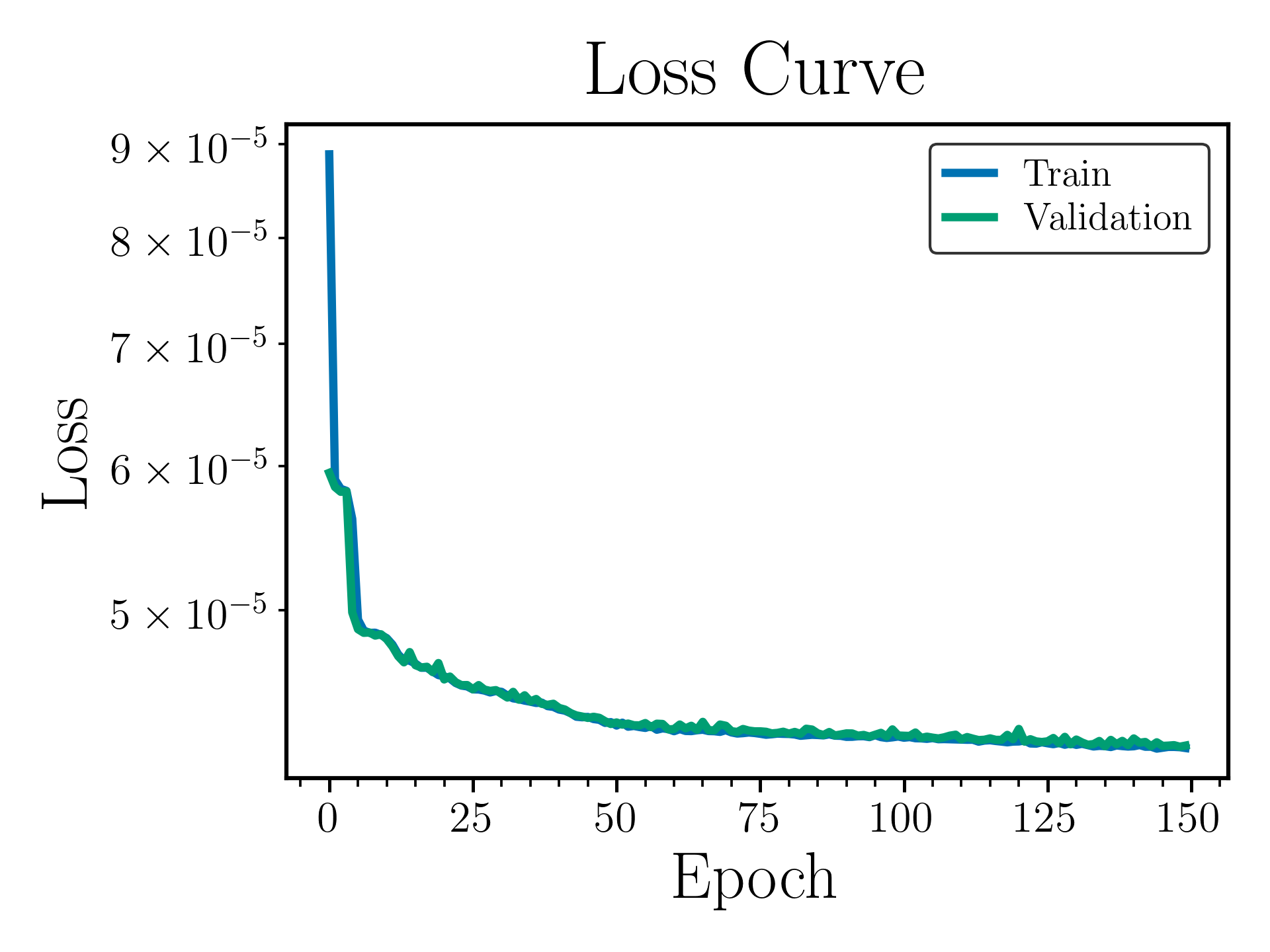}
    \caption{
        The loss function of the autoencoder during training.
    }
    \label{fig:loss-dynamics}
\end{figure}

\begin{figure}[!htb]
    \centering
    \includegraphics[width=0.8\textwidth]{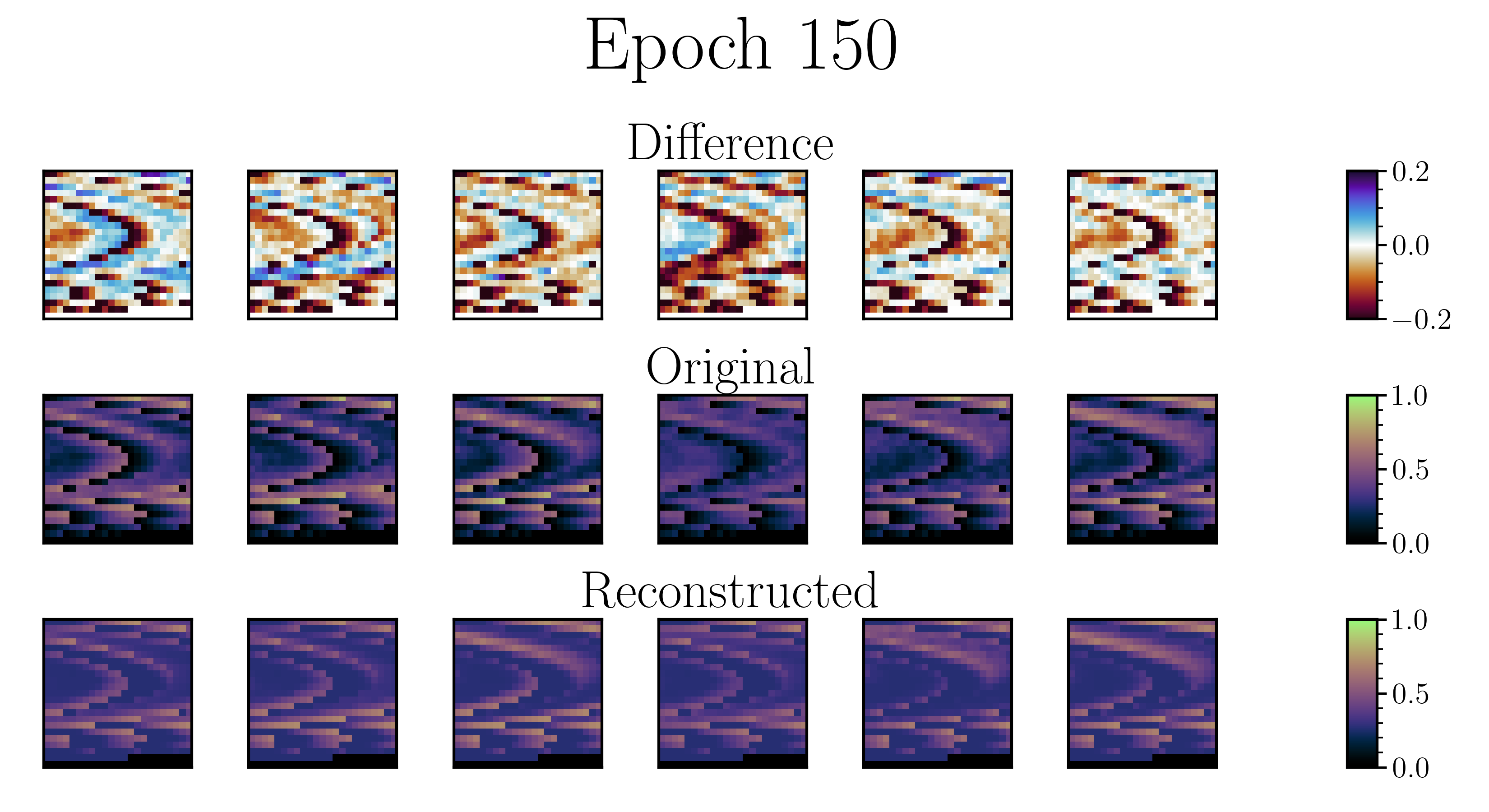}
    \caption{
        \textbf{Top panel:} Difference between the input and output of the AE for the pairwise distance between all backbone alpha carbons.
        \textbf{Middle panel:} Scaled input features to the AE.
        \textbf{Bottom panel:} Reconstructed features from the latent space.
    }
    \label{fig:loss}
\end{figure}
Figure \ref{fig:loss} provides a visual representation of the AE's performance on a set of 6 randomly selected conformations.
The top panel illustrates the difference between the scaled input data and the reconstructed output of the AE.
The middle panel displays the scaled input features to the AE, while the bottom panel presents the reconstructed features from the latent space, showcasing the capability of the AE to capture important features of the input data.

To gain insight into the relationships between the latent space features and relevant physical quantities influencing the conformational space, we analyzed the linear correlation between a few selected variables in Figure \ref{fig:correlation}.
The latent space features exhibit weak correlations with each other, implying that they capture distinct and complementary aspects of the system.
This observation suggests that the autoencoder effectively performs dimensionality reduction and representation learning, allowing it to extract relevant features while preserving important information about the system.
The moderate correlation with polymer radii of gyration is consistent with our previous single-chain studies, highlighting the significance of polymer size in determining calcium adsorption behavior.
The relatively weak correlation with interchain distance is attributed to the input data restriction of conformations within 1 $k_\mathrm{B} T$ of the minimum in the interchain PMF (0.70--1.07 nm), focusing on specific conformational states near the minimum.
\begin{figure}[!htb]
    \centering
    \includegraphics[width=0.5\textwidth]{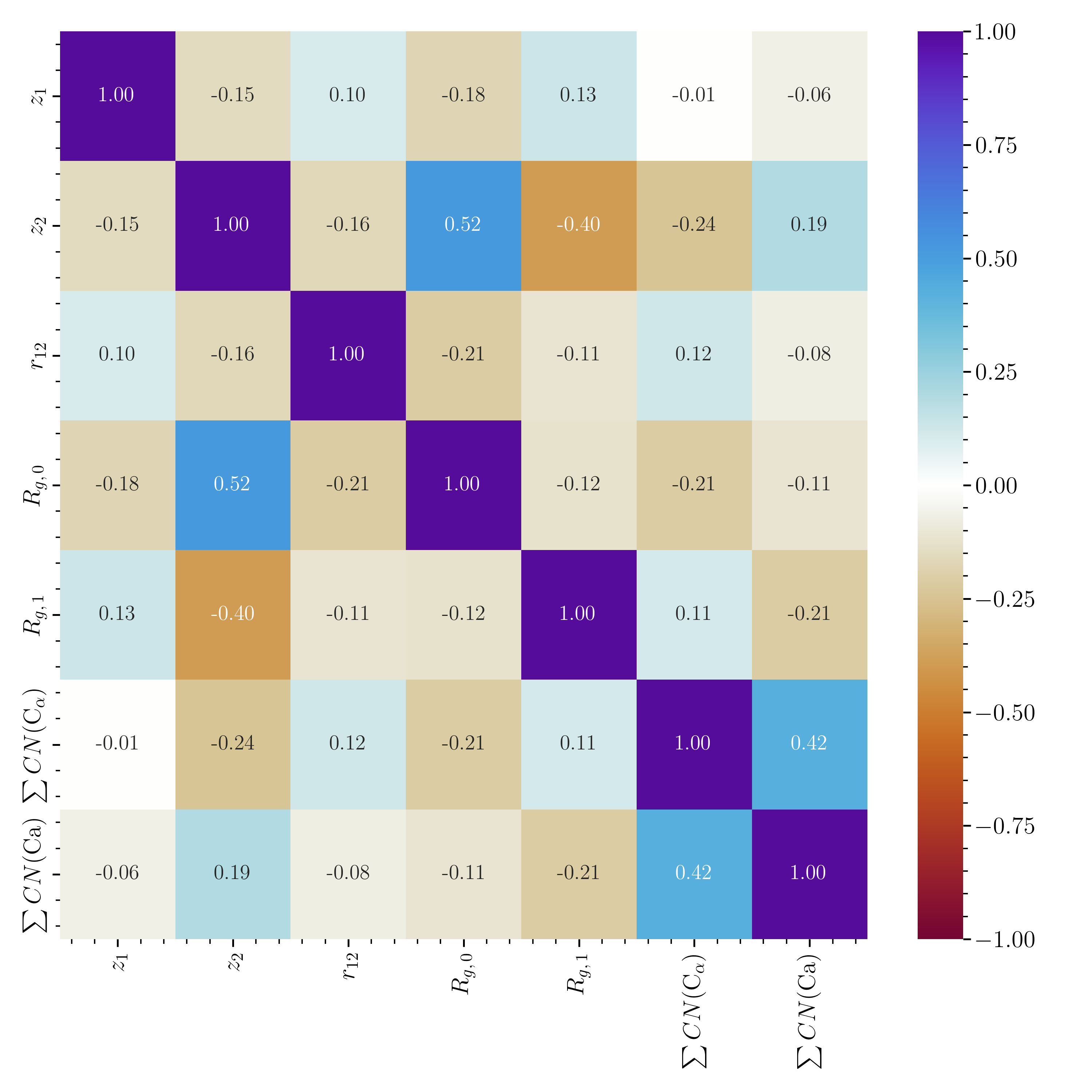}
    \caption{
        The correlation matrix of the latent space features $z_i$ with physical features of interest for the 16-mer PAA chains with 32 calcium ions within 1 $k_\mathrm{B} T$ of the minimum in the interchain PMF.
        The physical features are the interchain distance $r_{12}$, the chain radius of gyration $R_\mathrm{g}$, the total number of contacts between C$_\alpha$ atoms, and the total number of contacts between Ca$^{2+}$ ions and carboxylate carbons, respectively.
    }
    \label{fig:correlation}
\end{figure}

\endgroup

%
%